\documentclass[preprint,aps]{revtex4}
\usepackage[latin9]{luainputenc}
\usepackage{float}
\usepackage{amsmath}
\usepackage{color}
\usepackage{amssymb}
\usepackage{graphicx}
\usepackage{setspace}
\makeatletter

\@ifundefined{textcolor}{}
{%
 \definecolor{BLACK}{gray}{0}
 \definecolor{WHITE}{gray}{1}
 \definecolor{RED}{rgb}{1,0,0}
 \definecolor{GREEN}{rgb}{0,1,0}
 \definecolor{BLUE}{rgb}{0,0,1}
 \definecolor{CYAN}{cmyk}{1,0,0,0}
 \definecolor{MAGENTA}{cmyk}{0,1,0,0}
 \definecolor{YELLOW}{cmyk}{0,0,1,0}
}

\usepackage{amsfonts}
\usepackage{bm}
\usepackage[dvips]{epsfig}
\usepackage[dvips]{graphicx}
\usepackage{float}
\usepackage{dcolumn}
\usepackage{ifpdf}
\usepackage{multirow}

\newcommand{\sech}{\mathrm{sech}}
\newcommand{\be}{\begin{equation}}
\newcommand{\ee}{\end{equation}}
\newcommand{\bes}{\begin{subequations}}
\newcommand{\ees}{\end{subequations}}
\newcommand{\ben}{\begin{eqnarray}}
\newcommand{\een}{\end{eqnarray}}

\makeatother

\begin{document}

\title{Asymmetry engendered by symmetric kink-antikink scattering in a degenerate two-field  model}
\author{ Fabiano C. Simas$^{1,2}$,  K. Z. Nobrega$^3$, D. Bazeia$^{4}$, Adalto R. Gomes$^{1,5}$}
\email{fc.simas@ufma.br, kznobrega@ufc.br, bazeia@fisica.ufpb.br, argomes.gomes@ufma.br}

%\selectlanguage{english}

\affiliation{
$^1$ Programa de P\'os-Gradua\c c\~ao em F\'\i sica, Universidade Federal do Maranh\~ao\\Campus Universit\'ario do Bacanga, 65085-580, S\~ao Lu\'\i s, Maranh\~ao, Brazil
\\$^2$ Centro de Ci\^encias de Chapadinha-CCCh, Universidade Federal do Maranh\~ao\\65500-000, Chapadinha, Maranh\~ao, Brazil\\
$^3$ Departamento de Engenharia Teleinform\'atica, Universidade Federal do Cear\'a\\60455-640, Fortaleza, Cear\'a, Brazil\\
$^4$ Departamento de F\'isica, Universidade Federal da Para\'iba\\ 58051-970, Jo\~ao Pessoa, PB, Brazil\\
$^5$ Departamento de F\'isica, Universidade Federal do Maranh\~ao, Campus Universit\'ario do Bacanga, 65085-580, S\~ao Lu\'\i s, Maranh\~ao, Brazil.
}

\begin{abstract}

In this paper we analyze the scattering process in a two-field model in $(1+1)$-dimensions, with the special property to have several topological solutions: i) one with higher rest mass, characterized by a nested defect (lump inside a kink), and ii) four others having lower rest mass, degenerated, and characterized by a kink inside  kink.  We investigate  kink-antikink symmetric scattering, where the kink and antikink have higher rest mass and the same initial velocity modulus $v$.
The output of scattering  presents a wide range of behaviors, such as annihilation of the kink-antikink pair, the emission of radiation jets, the generation of oscillating pulses and the change of the topological sector. We show that the changing of the topological sector is favored, and only two of the four sectors are possible as outcomes. Moreover, despite the degeneracy in energy, the distribution of the final states is asymmetric in the phase space, being  an effect of the presence of vibrational states. 

%This is likely a general effect, present in other two-field models with phenomenological interest.

\end{abstract}

\maketitle

%%%%%%%%%%%%%%%%%%%%%%%%%%%%%%%%%%%%%%%%%%%%%%%%%%%%%%%

\section{ Introduction }

%%%%%%%%%%%%%%%%%%%%%%%%%%%%%%%%%%%%%%%%%%%%%%%%%%%%%%%

Solitary waves are  characterized by the very special property of free propagation without dispersion \cite{dau,wein,vacha}. Solitary waves of special interest are topological defects in nonlinear field theories, where stability of a localized energy density is related to a conserved topological current. The simplest topological defect is the $(1,1)$ dimensional kink (and the corresponding antikink), which appears in the model with degenerate minima potentials. The kink embedded in higher dimensions generate domain walls. Kinks and domain walls have been explored theoretically  in systems of variable complexity and energy scales. Of particular interest one can cite their importance for extended hadron model \cite{uchi} and the  description of the baryonic spectrum in low-energy effective action of bosonized two-dimensional QCD \cite{blas2, blas3}. Embedded in other dimensions, the kink generate domain walls, which can be generated following bubble collision acting as secondary gravitational wave sources \cite{dw1}, or as a possible description for dark matter \cite{dw2}. The collision of two colliding planar walls were used to describe the collision of nucleated bubbles considering the effects of small initial quantum fluctuations \cite{dw3}.

The symmetric kink-antikink scattering in nonintegrable single-field models has a  complex structure. For large initial velocity of the kink-antikink, one has the simple inelastic scattering with the pair colliding once and escaping to infinity. For small velocities one has the formation of a bion state that radiate continuously. Depending on the model one has for intermediate velocities, the possibility, for instance, of resonant bounce collision \cite{anninos}, oscillons \cite{hyp}, multiple kink-antikink pairs \cite{dsg}, formation of resonance windows by adding fermions \cite{azah1} and  spectral walls \cite{adam1}. That is, the scattering structure has an intricate pattern. The linear stability analysis of the defect gives eigenmodes and quasinormal modes, whose frequencies are a useful tool to get some understanding on the scattering \cite{camp1, dorey1, adalto2, dorey3,gani1}. A more recent approach considers moduli space approximation \cite{ms1, ms2}  to understand the scattering dynamics. 

 In systems with two scalar fields, we can expect an even more intricate behavior. Indeed, the occurrence of kink solutions with internal structure is favored by the presence of two scalar fields in some supersymmetric theories \cite{shif}.
Kink-antikink dynamics in models with two scalar fields was investigated for instance in the Refs. \cite{alonso4,alonso5,alonso6,alonso7,roman1}.  In the Ref. \cite{adam2} the existence of the spectral wall phenomenon in models with multiple scalar fields was confirmed.  In late cosmology two-field models were used to investigate the effects of the interaction between dark matter and dark energy \cite{berto, grigo}. 
In inflationary cosmology two scalar fields are useful to unify inflation in the early universe. That is, one scalar field can explain dark energy and the other scalar field can explain dark matter \cite{kazu}. 

Of particular interest in the present work is the model of two coupled scalar fields $\phi,\chi$ in $(1,1)$ dimensions introduced in the Ref. \cite{bazeia4}. The model has a coupling parameter $r$ such that in a topological sector there are explicit kink ($K_{21}$) and antikink ($\bar K_{21}$) solutions for $0<r<1/2$ in a form of nested defects, where the field $\phi$ has a kink (antikink) profile whereas the $\chi$ field has a lump one. There are four other topological sectors with kinks degenerated with a lower energy, where both fields have a kink (or antikink) profile. The formed defects have internal structure similar to the obtained in the Bloch wall scenario obtained in the Ginzburg-Landau equation describing magnetic systems \cite{mot1}. The presence of an internal structure, caused by the introduction of a new scalar field, is in charge of controlling the domain wall thickness. The collision process with two or more scalar fields can then demonstrate how to describe this new degree of freedom. Also this model was considered in a scenario with domain wall with internal structure embedded in other defect of higher dimensions \cite{brito}.

In the next section we present the model. In the section \ref{sec3} we present  the numerical analysis of kink-antikink scattering, showing how the initial higher energy solution can be traded by other solutions degenerate in energy, with some interesting effects. We conclude in the section \ref{sec4}.

%%%%%%%%%%%%%%%%%%%%%%%%%%%%%%%%%%%%%%%%%%%%%%%%%%%%%%%%%%%%%%%%%%%%%%%%%%%

\section { The Model } \label{sec2}

%%%%%%%%%%%%%%%%%%%%%%%%%%%%%%%%%%%%%%%%%%%%%%%%%%%%%%%%%%%%%%%%%%%%%%%%%%%

We consider a two-coupled scalar field model with $(1,1)$-dimensional governed by the action 
\be
S=\int dt dx \bigg[ \frac12 \partial_{\mu} \phi \partial^{\mu} \phi +\frac12 \partial_{\mu} \chi \partial^{\mu} \chi   - V(\phi, \chi)  \bigg],
\label{action}
\ee
where the potential $V=V(\phi,\chi)$ is a function of partial derivatives of a smooth function $W(\phi,\chi)$ as
\be
V(\phi,\chi)=\frac12 W_\phi^2 + \frac12 W_\chi^2.
\ee
The energy density of static solutions are given by
\be
\mathcal{E} (x) = \pm \frac{dW}{dx} + \frac 12 \biggl( \frac{d\phi}{dx} \mp W_\phi \biggr)^2 + \frac 12 \biggl( \frac{d\chi}{dx} \mp W_\chi \biggr)^2.
\ee
Then, solutions satisfying the first-order equations
\be
 \frac{d\phi}{dx} = \pm W_\phi, \,\,\,  \frac{d\chi}{dx} = \pm W_\chi,
\label{1ord}
\ee
are BPS solutions and minimize energy. The plus sign in the Eqs. (\ref{1ord}) will result in kinks whereas the minus sign to antikinks. The energy of the solutions are given by $E_{BPS} = |W[ \phi(+\infty), \chi(+\infty)] -  W[ \phi(-\infty), \chi(-\infty)]|$.
In this work we consider the function \cite{bazeia4}
\be
W(\phi,\chi) = \phi - \frac13 \phi^3 - r \phi \chi^2,
\ee
with $r$ a positive constant. This corresponds to the potential
\be
V(\phi, \chi) = \frac12 (1-\phi^2-r\chi^2)^2 + 2r^2\phi^2\chi^2,
\label{pot}
\ee
 This potential has minima at $v_{1,2}(\pm 1,0)$ and $v_{3,4}(0,\pm 1/\sqrt{r})$, and have five BPS sectors connecting the minima $v_i$ and $v_j$ with energy $E_{ij}$:
one with energy $E_{12}= 4/3$ and four degenerate sectors with energy $E_{13}=E_{14}=E_{23}=E_{24}=2/3$. 

The equations of motion are given by
\begin{eqnarray}
\frac{\partial^2 \phi}{\partial t^2}-\frac{\partial^2 \phi}{\partial x^2}+\frac{dV(\phi,\chi)}{d\phi} & = & 0,
\label{eqm1}\\
\frac{\partial^2 \chi}{\partial t^2}-\frac{\partial^2 \chi}{\partial x^2}+\frac{dV(\phi,\chi)}{d\chi} & = & 0.
\label{eqm2}
\end{eqnarray}
and the first-order equations for kinks are given by
\begin{eqnarray}
\frac{d\phi}{dx} &=& 1 - \phi^2 - r\chi^2, \\
\frac{d\chi}{dx} &=& -2 r \phi \chi. 
\end{eqnarray}
These equations can be solved using the trial orbits method \cite{bazeia1} or after finding an integrating factor \cite{,alonso7,guill}, leading to the $K_{21}-$kink solutions connecting the minima $v_2(-1,0)$ and $v_1(1,0)$ for $0<r<1/2$, and  given by  
\begin{eqnarray}
\phi_{21}(x,r) &=& \tanh(2 r x), \label{kink-K21} \\
\chi_{21}(x,r) &=&  \sqrt{\frac1r -2}\sech(2 r x). \label{lumpK21}
\end{eqnarray}
 Similarly, $\bar K_{21}-$antikink solutions of Eqs. (\ref{1ord}) with minus sign connecting the minima $v_2(1,0)$ and $v_1(-1,0)$ for $0<r<1/2$ are given by  
\begin{eqnarray}
\bar\phi_{21}(x,r) &=& -\phi_{21}(x,r) = -\tanh(2 r x), \label{antikink-K21} \\
\bar\chi_{21}(x,r) &=& \chi_{21}(x,r) =  \sqrt{\frac1r -2}\sech(2 r x). \label{antilumpK21}
\end{eqnarray}

The other degenerated solutions for the other topological sectors are i) The $K_{31}-$kink solution connecting the minima $v_3(0, 1/\sqrt r)$ and $v_1(1, 0)$; The $K_{41}-$kink solution connecting the minima $v_4(0, -1/\sqrt r)$ and $v_1(1, 0)$;  iii) The $K_{23}-$kink solution connecting the minima $v_2(-1, 0)$ and $v_3(0, 1/\sqrt r)$; iv)  iii) The $K_{24}-$kink solution connecting the minima $v_2(-1, 0)$ and $v_4(0, -1/\sqrt( r)$. These solutions are degenerate in energy and orbits in the phase space $(\phi,\chi)$ and be found using the integrating factor \cite{bw}. However, explicit $x$-dependence of them can be found only for very specific values of $r$.  

%%%%%%%%%%%%%%%%%%%%%%%%%%%%%%%%%%%%%%%%%%%%%%%%%%%%%%%%%%%%%%%%%%%%%
\begin{figure}
	\includegraphics[{angle=0,width=8cm,height=6cm}]{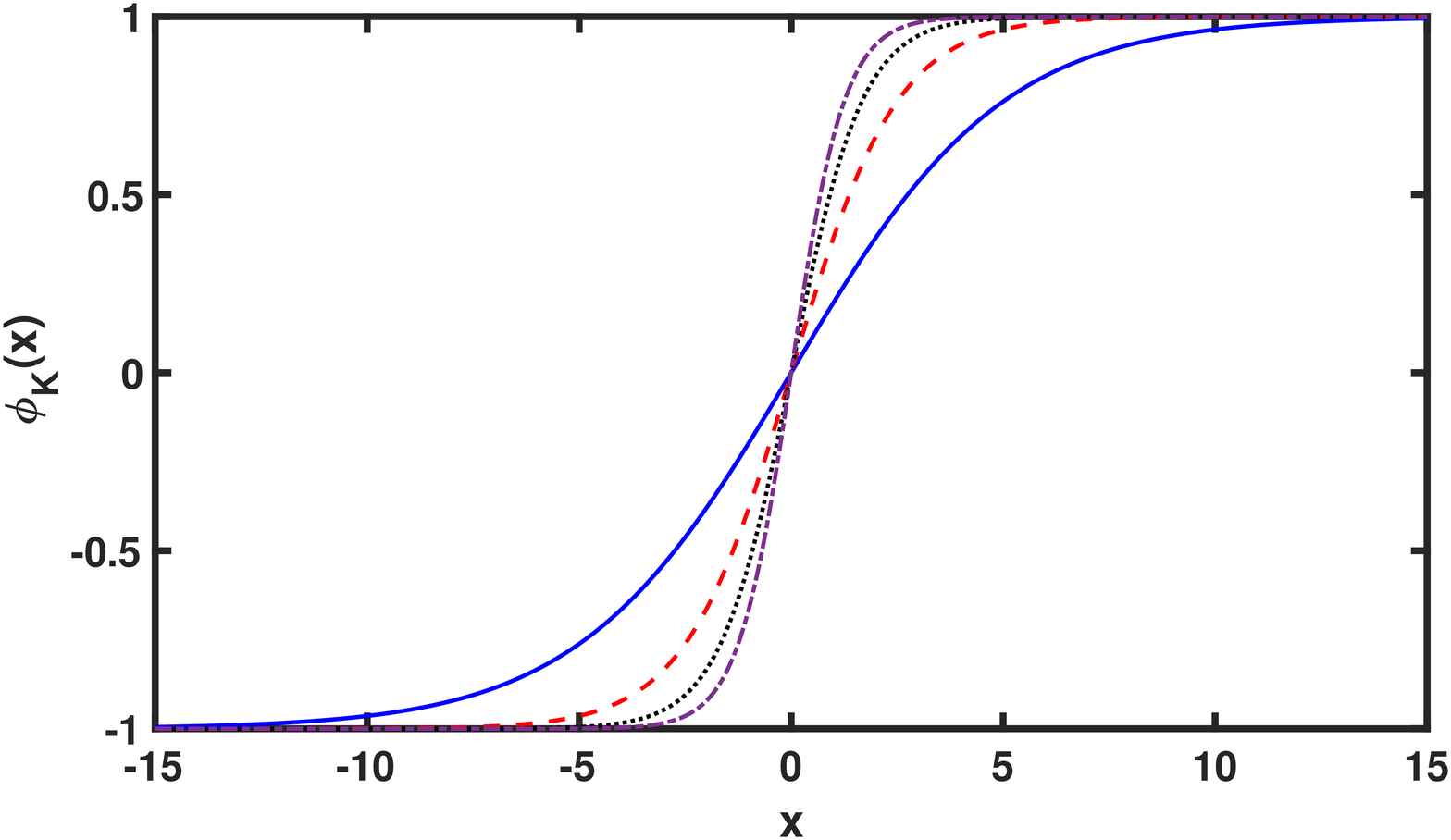}
	\includegraphics[{angle=0,width=8cm,height=6cm}]{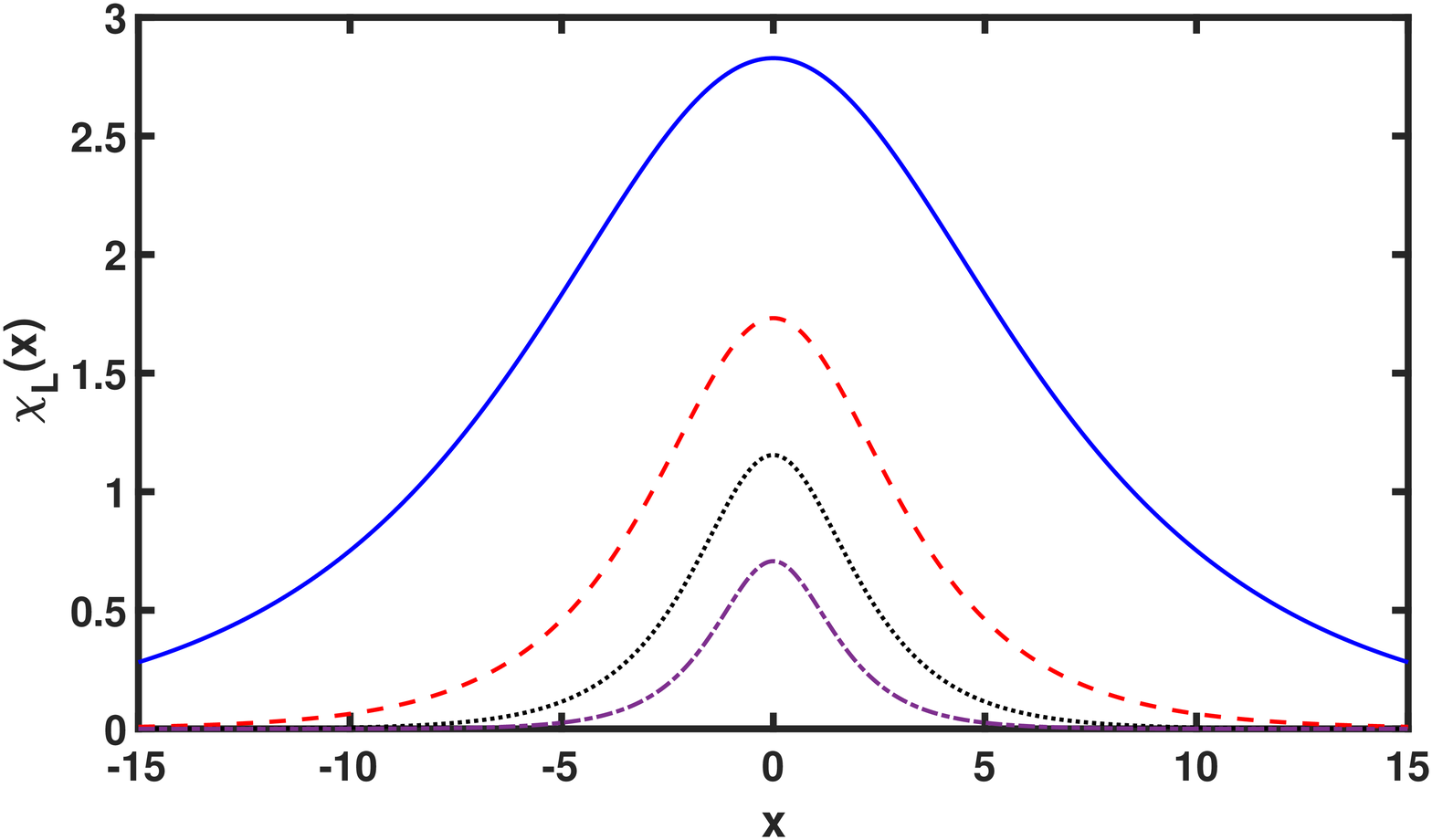}
\includegraphics[{angle=0,width=8cm,height=6cm}]{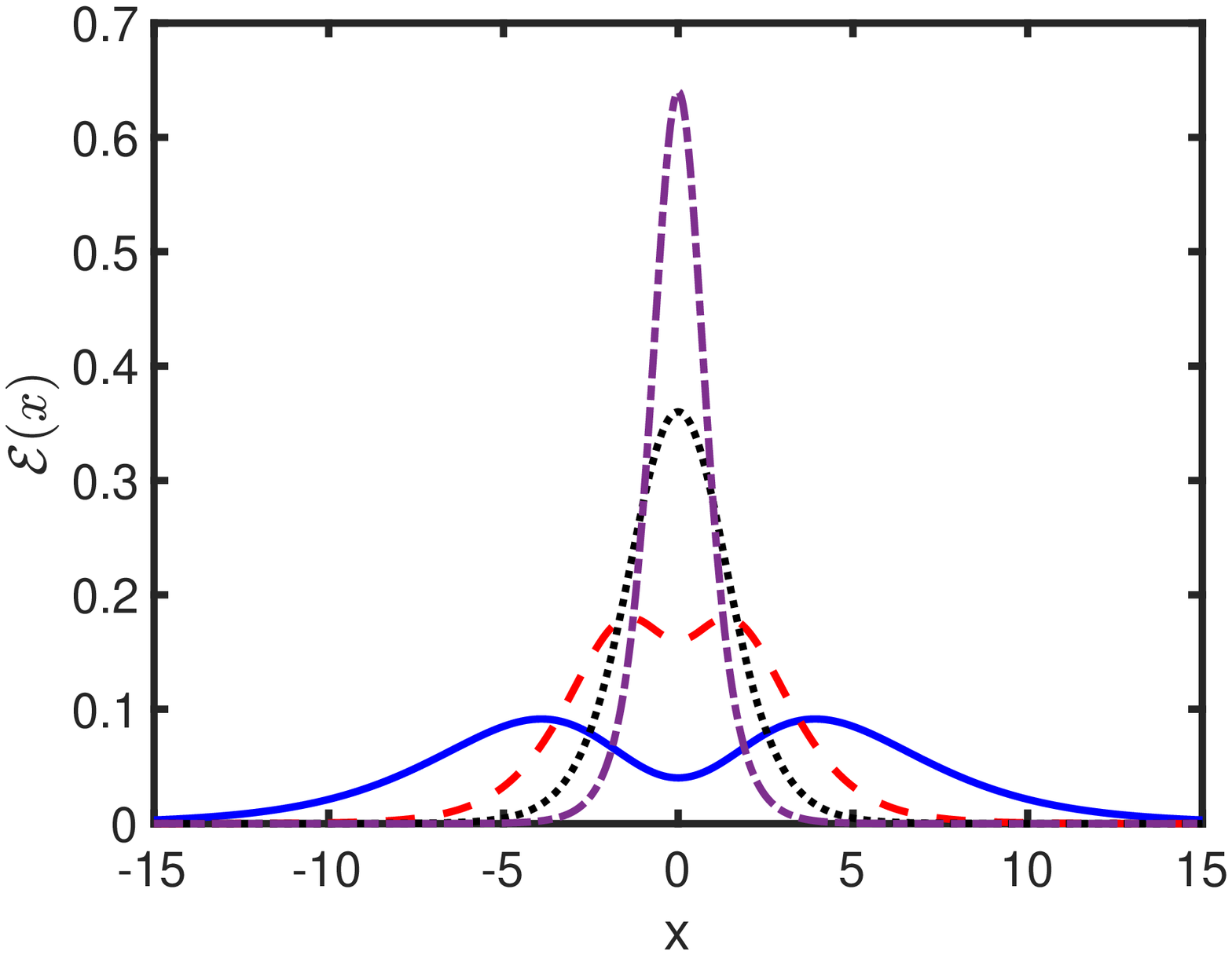}
	\caption{The $K_{21}-$kink solution connecting the minima $v_2(-1,0)$ and $v_1(1,0)$, showing the components a) $\phi_{21}(x)$, b) $\chi_{21}(x)$ and (c) the energy density $\mathcal{E} (x)$. In the figures we fixed $r=0.1$ (blue line), $r=0.2$ (red dash), $r=0.3$ (black dot) and $r=0.4$ (purple dash-dot).}
	\label{sol}
\end{figure}
%%%%%%%%%%%%%%%%%%%%%%%%%%%%%%%%%%%%%%%%%%%%%%%%%%%%%%%%%%%%%%%%%%%%%

Note also that in all degenerate solutions, both fields $\phi$ and $\chi$ interpolate between different minima, whereas for the solution $K_{21}$, the field $\chi$ has a lump structure. This is depicted in the Figs. \ref{sol}a and \ref{sol}b. The energy density $\mathcal{E} (x)$ of this solution is depicted in the Figs. \ref{sol}c for several values of $r$. Note from the figure that for $0.3<r<0.5$ one has a peak centered at $x=0$. For $r=0.3$ there is a plateau at $x=0$. For $0<r<0.3$ there is the appearance of two peaks in the energy density, showing that the defect has an internal structure. That is, we have a nested defect, where the field $\chi$ is in the core of the defect and contributes to enrich its energy density. Domain walls with internal structure were observed in ferromagnets \cite {hans}. A similar solution was obtained for an extension of this model with extra dimensions and gravity in the Ref. \cite{bazeia3}. Since this solution is more complex and more energetic, with known analytical solution for the range $0<r<1/2$, in the following section we will consider the kink-antikink scattering with solutions $K_{21}$ and $\bar K_{21}$. We remark that the Ref. \cite{alonso7} already studied kink-antikink scattering in this model, but restricted to $r=1$ and for different solutions, whereas in the present work we consider $0<r<1/2$.

%%%%%%%%%%%%%%%%%%%%%%%%%%%%%%%%%%%%%%%%%%%%%%%%%%%%%%%%%%%%%%%%%%%%%

\section{Numerical results}\label{sec3}

%%%%%%%%%%%%%%%%%%%%%%%%%%%%%%%%%%%%%%%%%%%%%%%%%%%%%%%%%%%%%%%%%%%%%

In this section we describe the numerical results concerning the scattering process of the $K_{21}-$kink. For this process, we solved the two equations of motion (Eqs. (\ref{eqm1}) and (\ref{eqm2}))  with a $4^{th}$ order finite-difference method with a spatial step $\delta x=0.05$. We fixed $x=\pm x_0=\pm 20$ for the initial symmetric position of the pair. For the time dependence we used $6^{th}$ order sympletic integrator method, with a time step $\delta t=0.02$. We used the following initial conditions for scattering

\begin{eqnarray}
\phi(x,0,x_0,v,r)&=&\phi_K(x+x_0,0,v,r) - \phi_K(x-x_0,0,-v,r)-1\\
\dot{\phi}(x,0, x_0, v,r)&=&\dot{\phi_K}(x+x_0,0,v,r) - \dot{\phi_K}(x-x_0,0,-v,r),
\end{eqnarray} 
and
\begin{eqnarray}
\chi(x,0, x_0, v,r)&=&\chi_L(x+x_0,0,v,r) + \chi_L(x-x_0,0,-v,r)\\
\dot{\chi}(x,0, x_0, v,r)&=&\dot{\chi_L}(x+x_0,0,v,r) - \dot{\chi_L}(x-x_0,0,-v,r),
\end{eqnarray} 
where $\phi_K(x,t,v,r)=\phi_{21}(\gamma(x-vt),r)$ and $\chi_L(x,t,v,r)=\chi_{21}(\gamma(x-vt),r)$ means a boost for the static solution with $\gamma=(1-v^2)^{-1/2}$.

%%%%%%%%%%%%%%%%%%%%%%%%%%%%%%%%%%%%%%%%%%%%%%%%%%%%%%%%%%%%%%%%%%%%%
\begin{figure}
	\includegraphics[{angle=0,width=14cm, height=8cm}]{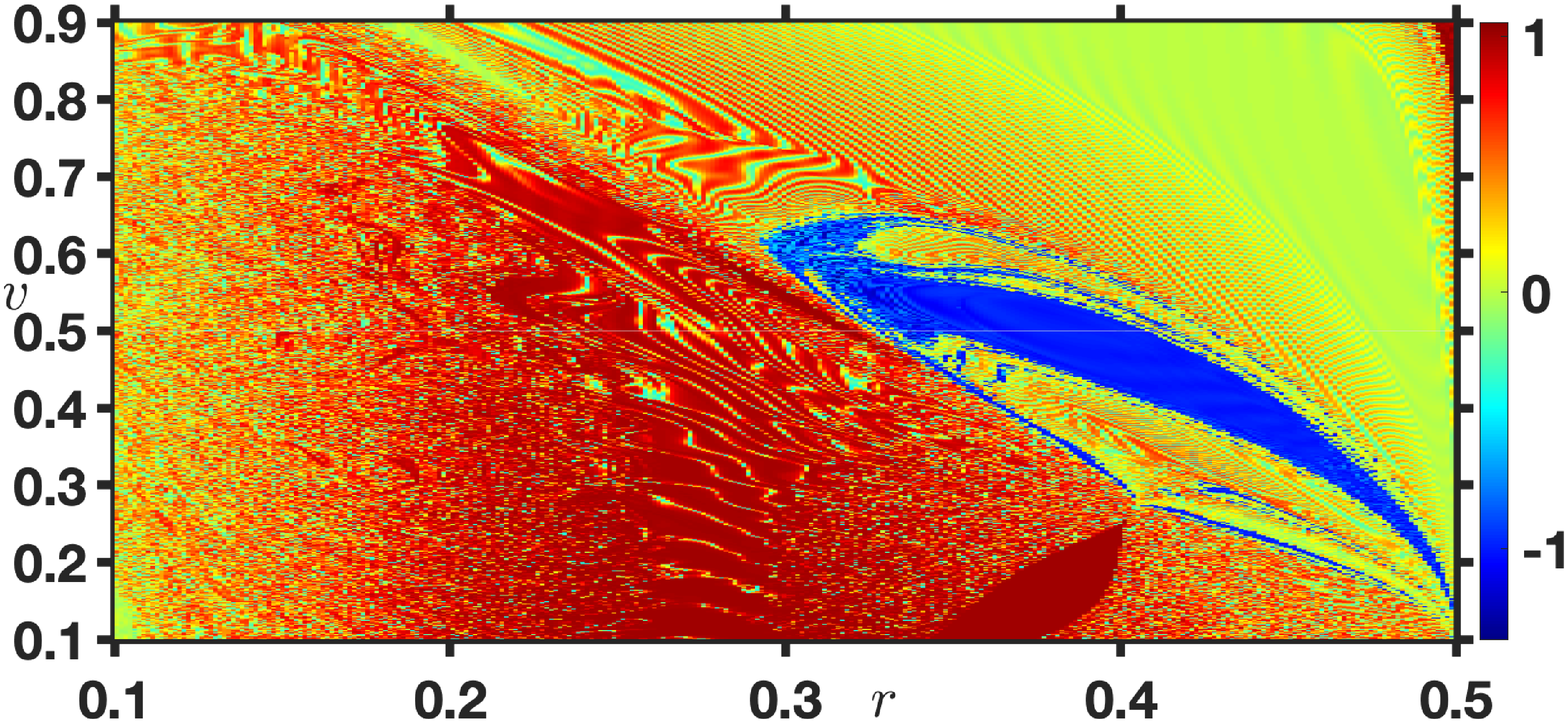}\\
\includegraphics[{angle=0,width=14cm, height=8cm}]{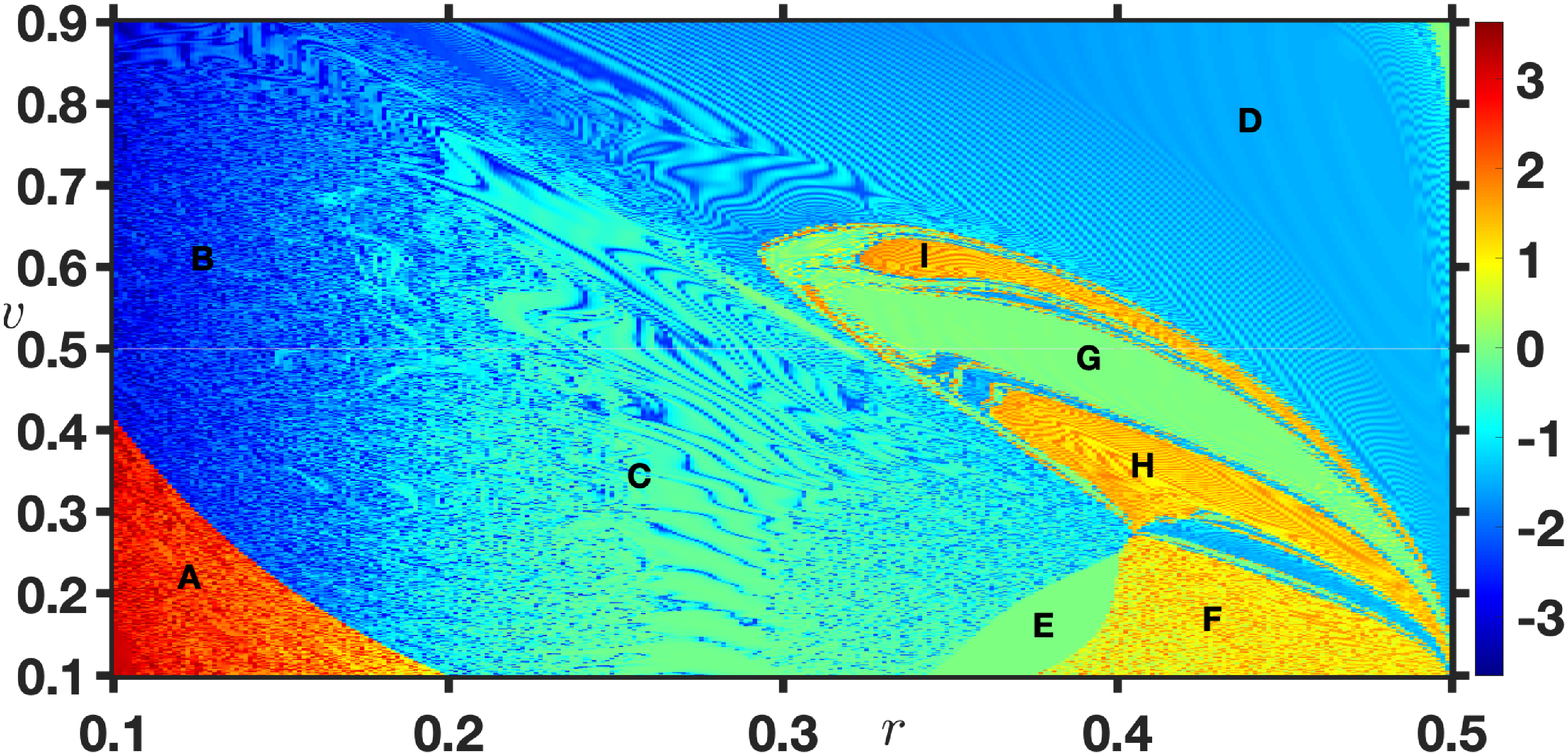} \\%,height=10cm
	\caption{Output of the $K_{21}\bar K_{21}$ collision process: a) (top) Final state of the scalar field $\phi_K(x=0,t_f)$, b) (bottom) Final state of the scalar field $\chi_K(x=0,t_f)$. Here one can also see the different regions labeled from A to I. }
	\label{mosaic}
\end{figure}
%%%%%%%%%%%%%%%%%%%%%%%%%%%%%%%%%%%%%%%%%%%%%%%%%%%%%%%%%%%%%%%%%%%%%

To better understand the scattering process of the $K_{21}-$kink, we will consider separately its components $\phi(x)$ and $\chi(x)$.  The structure of our results of scattering process is presented in the Figs. \ref{mosaic}a and \ref{mosaic}b.  There we observe the bidimensional $(v,r)$ phase space, that corresponds the final state of scalar fields $\phi(x=0,t_f)$ and $\chi(x=0,t_f)$. We noticed in the figures the presence of intricate patterns. It is important to note that the phase space structure is roughly the same in both figures. This indicate that the scalar fields $\phi$ and $\chi$ do not decouple after scattering, and we have still a defect  with internal structure. From the the diagrams we can identify nine regions, labeled from A to I, and showed in the Fig. \ref{mosaic}b. Now we will consider separately the characteristics of each region. Remember that we are considering collisions of the type 
$K_{21}\bar K_{21}$, where the solution $K_{21}$ interpolates between the vacuum $v_2(-1,0)$ and $v_1(1,0)$. In particular, the behavior for $x\to -\infty$ is fixed. This means that only two of the four topological sectors are possible as outputs.
%%%%%%%%%%%%%%%%%%%%%%%%%%%%%%%%%%%%%%%%%%%%%%%%%%%%%%%%%%%%%%%%%%%%%
\begin{figure}
	\includegraphics[{angle=0,width=4cm}]{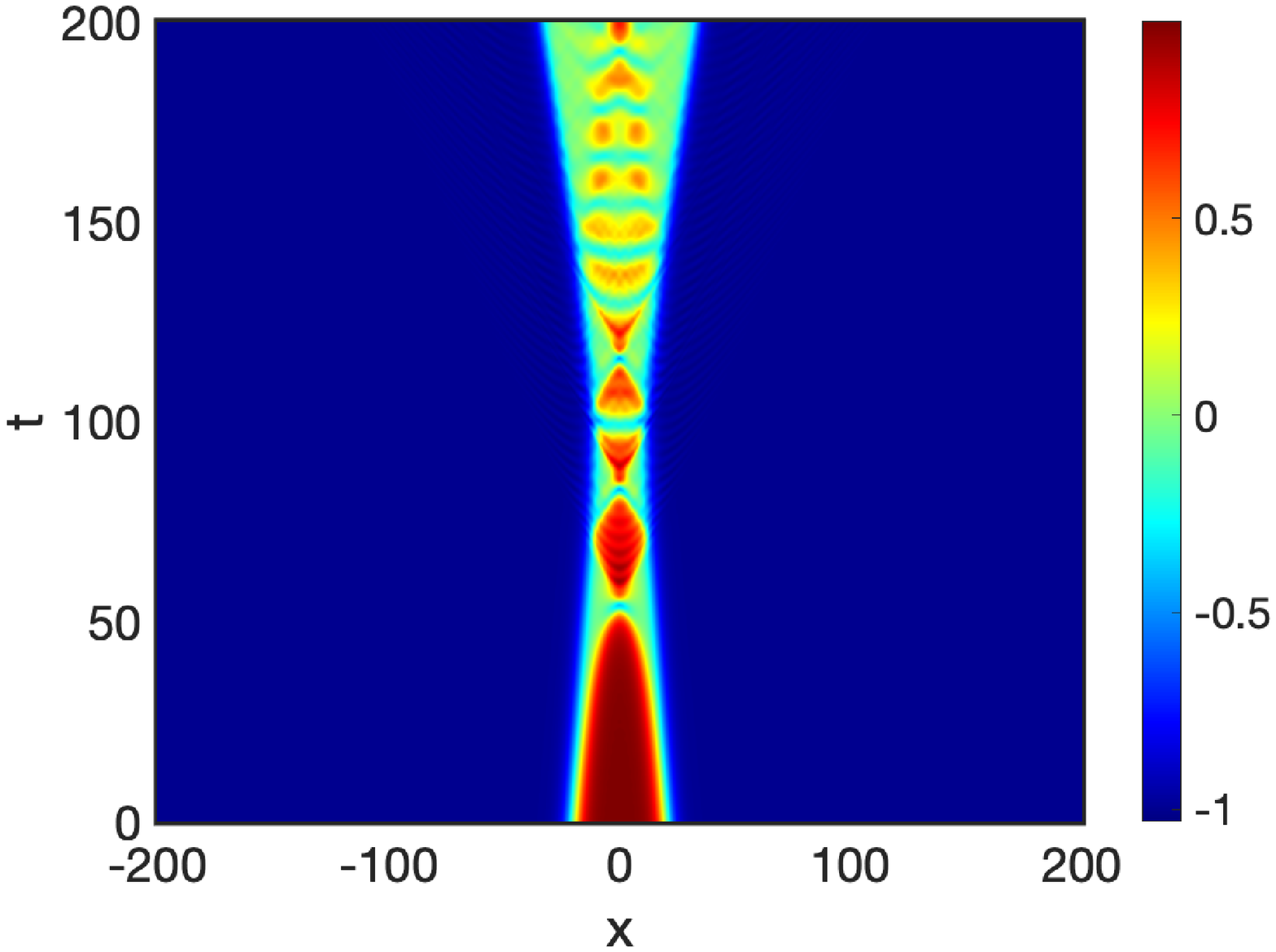}
	\includegraphics[{angle=0,width=4cm}]{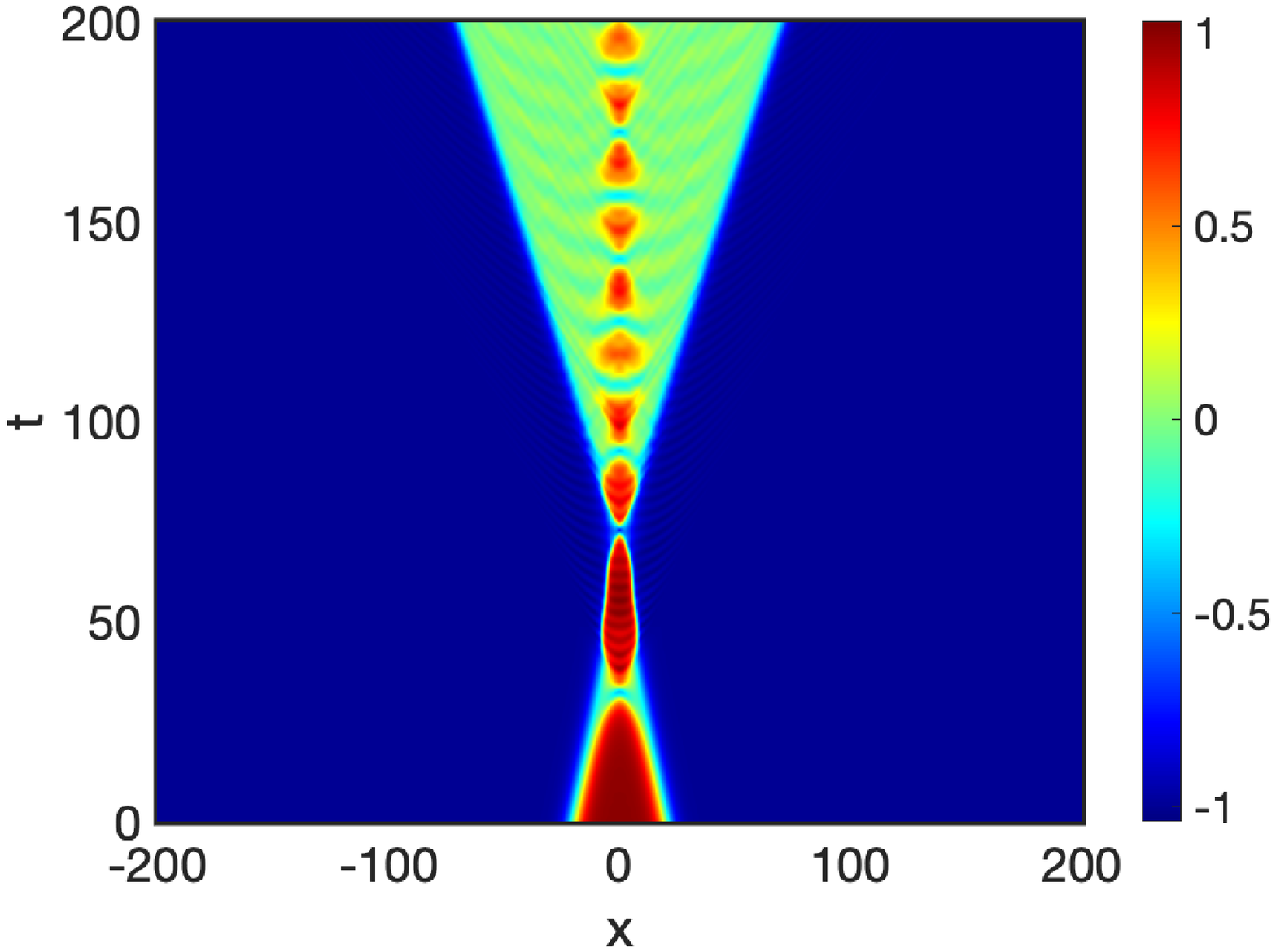}
	\includegraphics[{angle=0,width=4cm}]{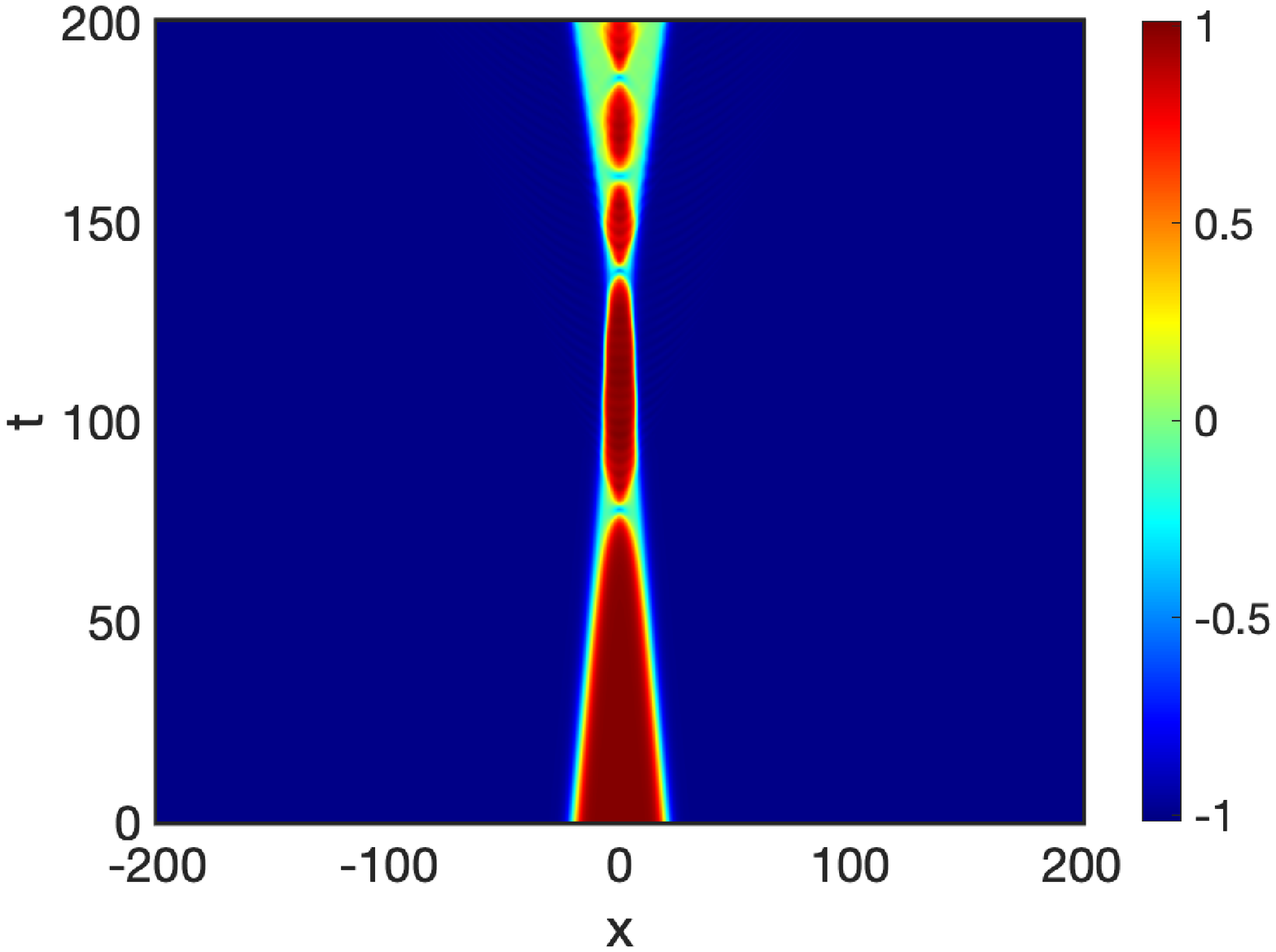}
	\includegraphics[{angle=0,width=4cm}]{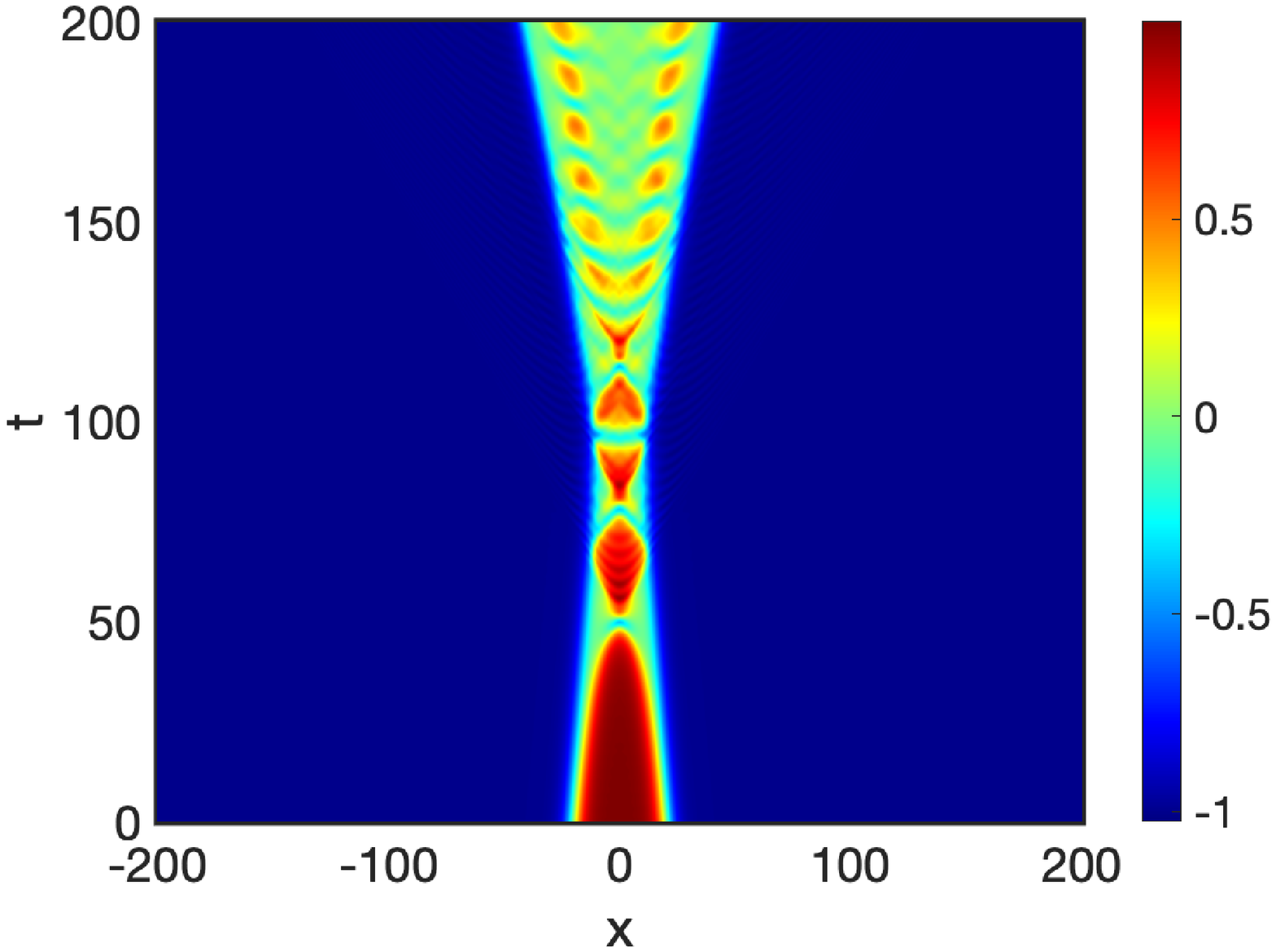}\\
\includegraphics[{angle=0,width=4cm}]{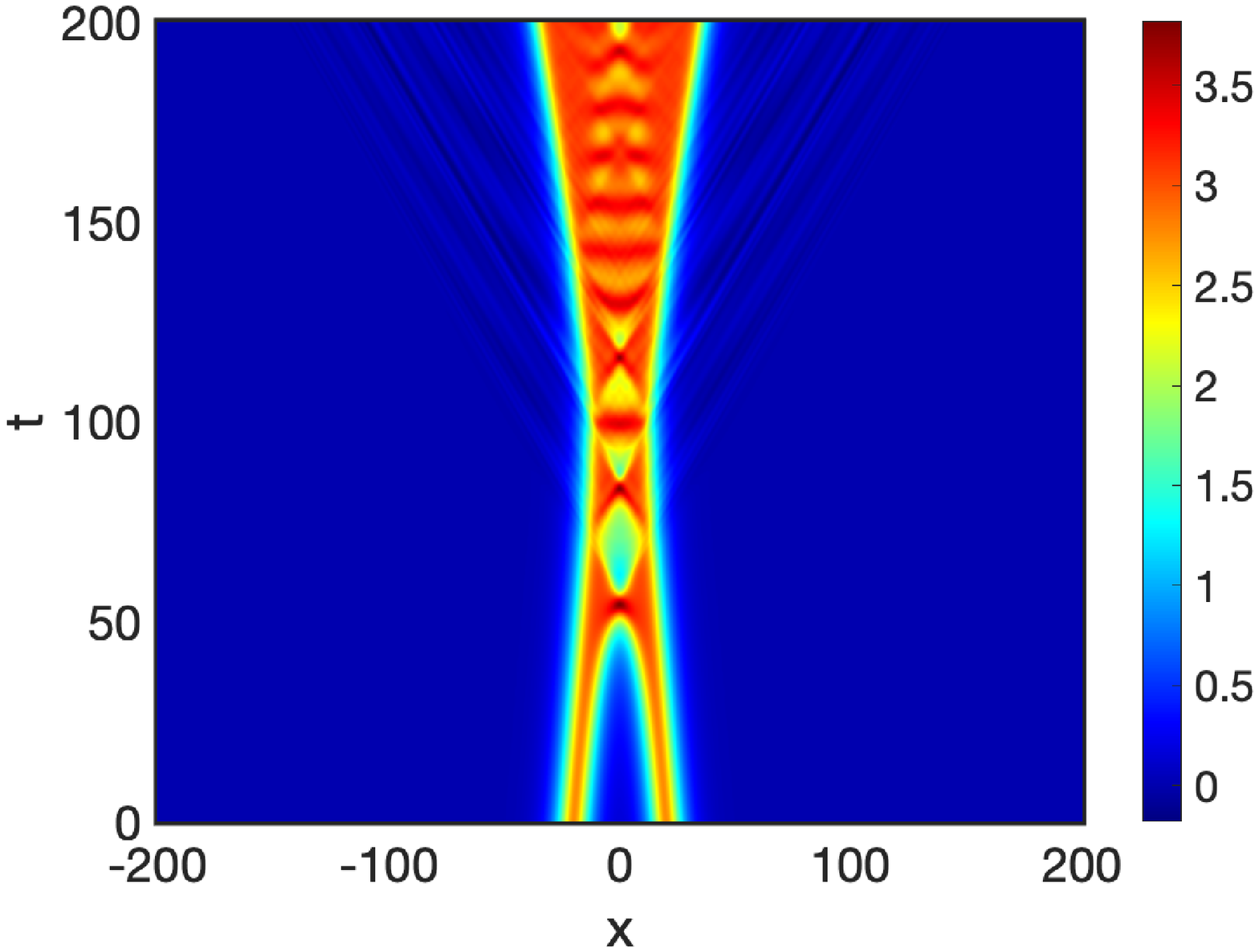}
	\includegraphics[{angle=0,width=4cm}]{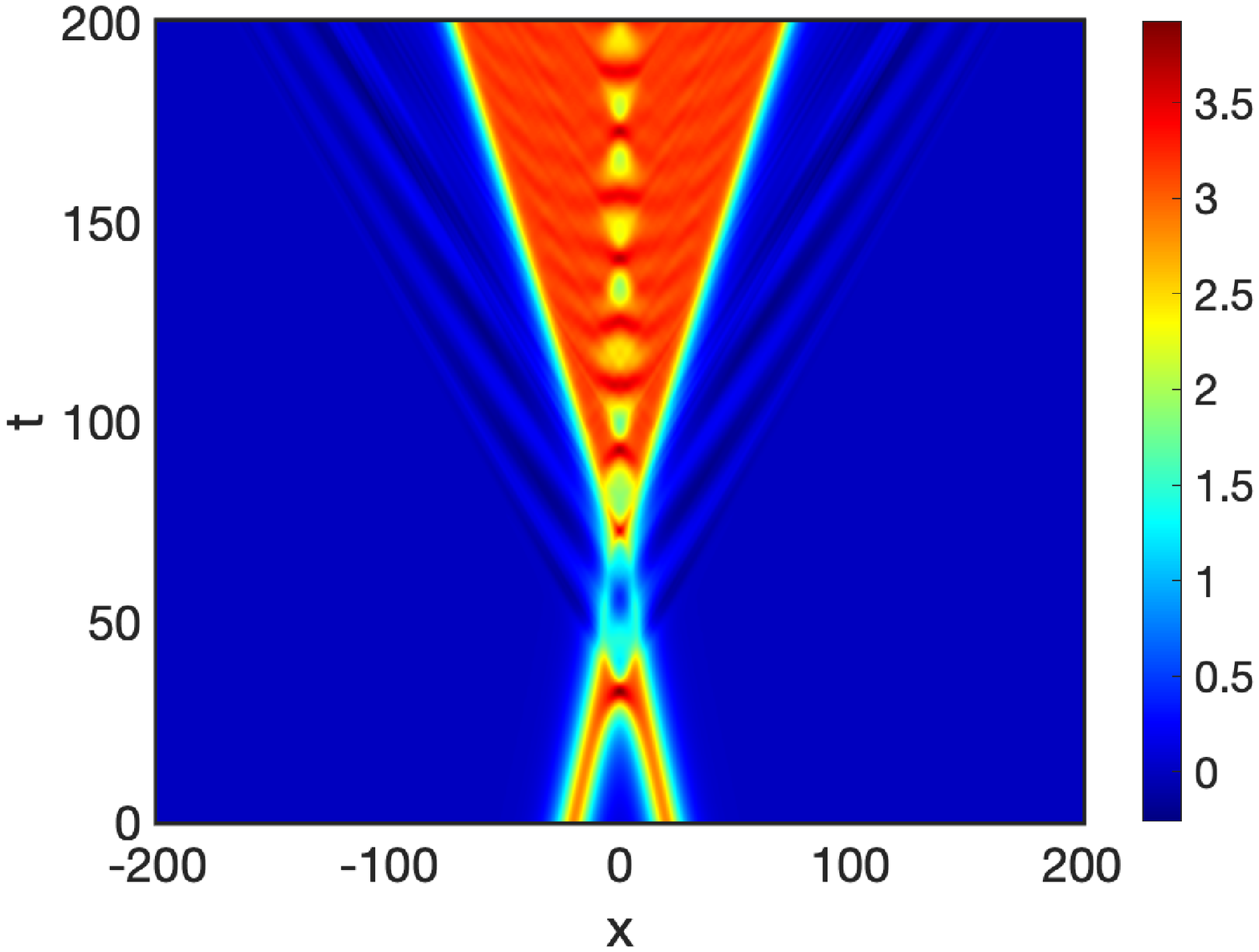}
	\includegraphics[{angle=0,width=4cm}]{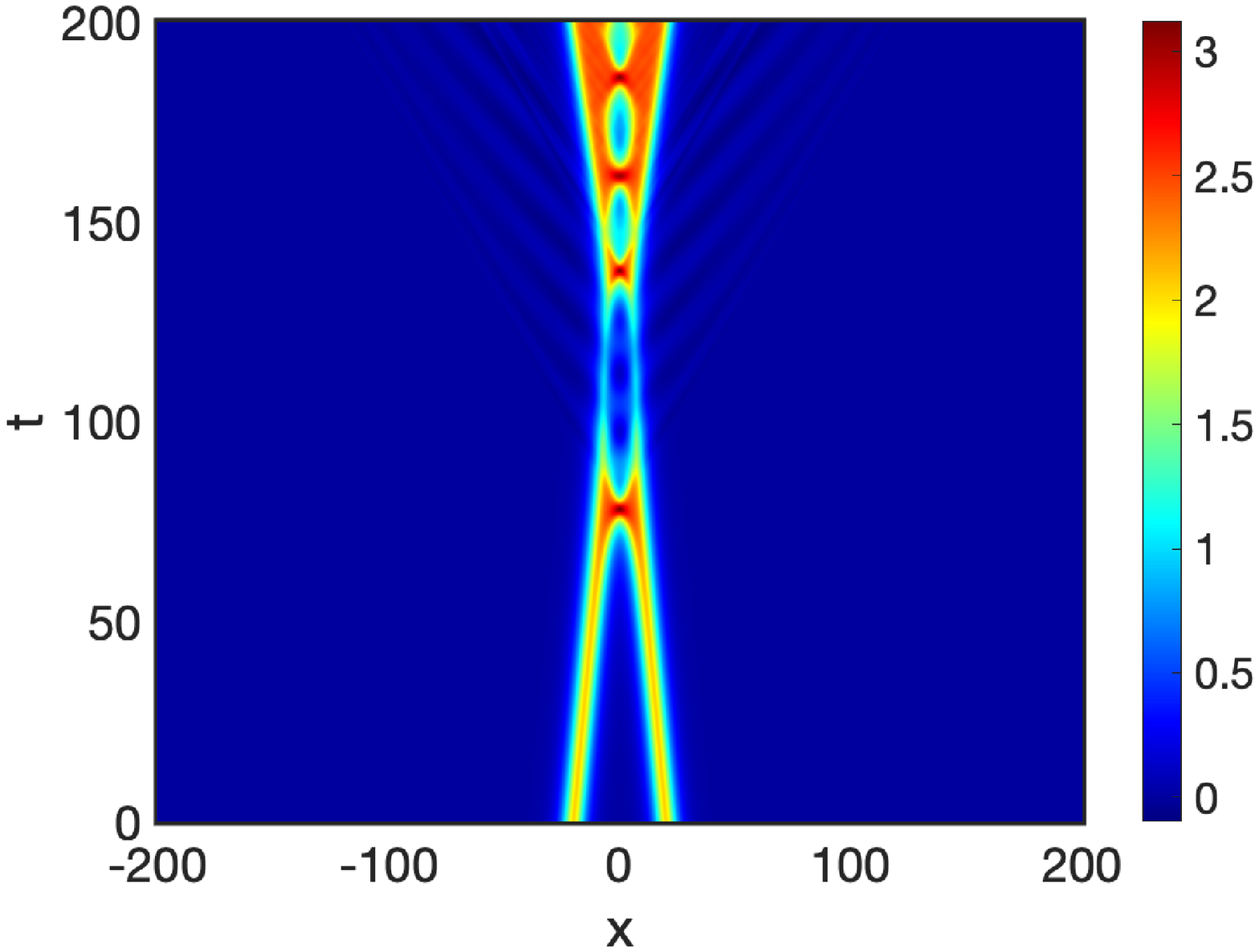}
	\includegraphics[{angle=0,width=4cm}]{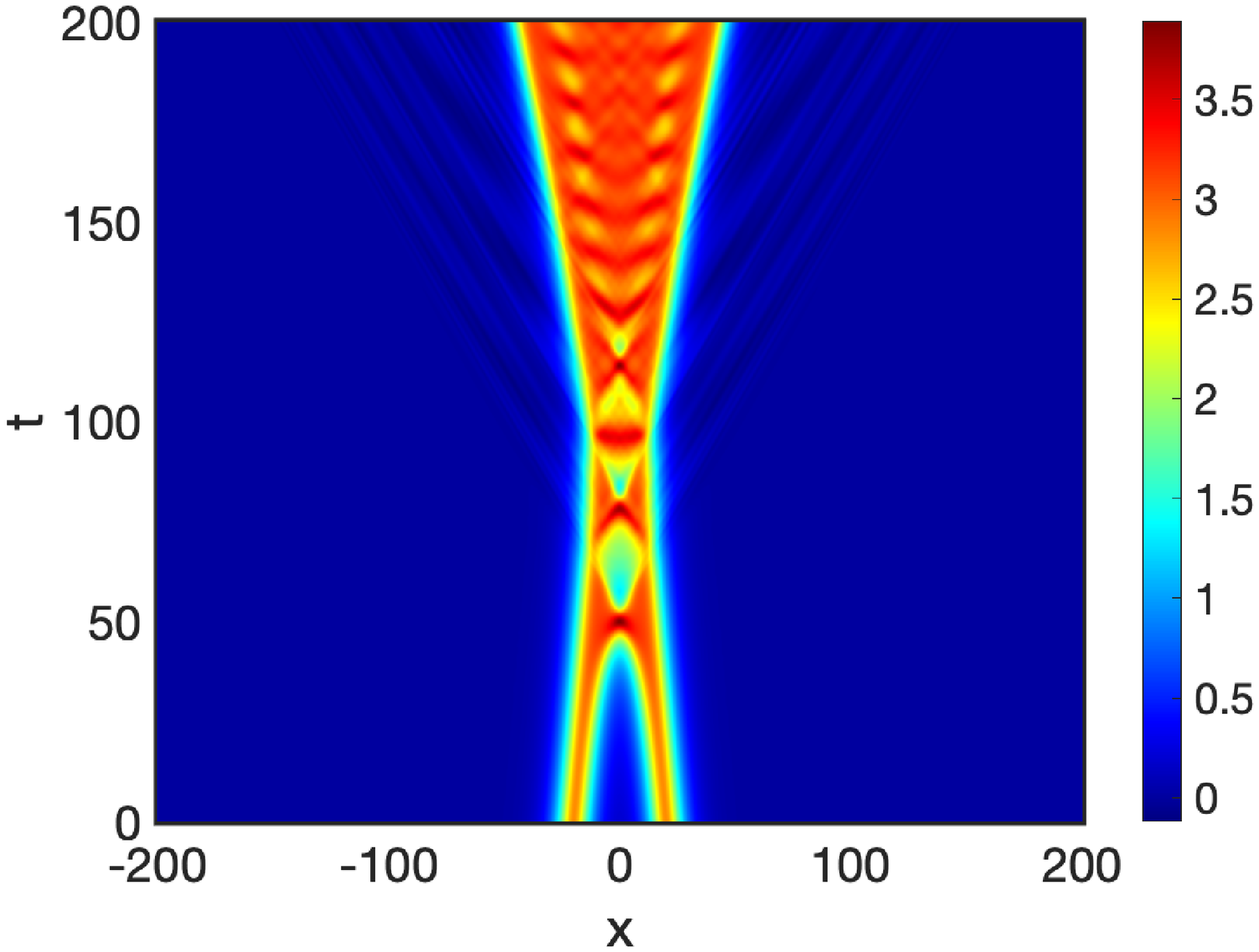}
	\caption{Region A - $\phi$-component (top)  and $\chi$-component (bottom) for (a) $v=0.14$ with $r=0.104$ (b) $v=0.37$ with $r=0.10$, (c) $v=0.158$ with $r=0.162$ and (d) $v=0.15$ with $r=0.10$.}
	\label{colA}
\end{figure}
%%%%%%%%%%%%%%%%%%%%%%%%%%%%%%%%%%%%%%%%%%%%%%%%%%%%%%%%%%%%%%%%%%%%%

The region A is characterized for small values of $r$ and $v$. This region has collisions represented in the Figs. \ref{colA}. For $x<0$ note that the $\phi$-component changes from $-1 \to 1$ to $-1 \to 0$, whereas the $\chi$-component changes to $0 \to 3 \to 0$ to $0 \to 3$, meaning that the $K_{21}$-kink changes to $K_{23}$ after scattering. For $x>0$ one can make a similar reasoning: the $\phi$-component changes from $1 \to -1$ to $0 \to  -1$, and the $\chi$-component from  $0 \to 3 \to 0$  to $3 \to 0$, meaning that the $\bar K_{21}$-antikink changes to $\bar K_{23}$ after scattering. Then, the collision can be characterized by $K_{21}+\bar K_{21} \to K_{23} + \bar K_{23}$ and the production of a stationary oscillation around $x=0$. We noted that the emission of radiation is more evidently produced by the $\chi$-component. This region are more complex because it coincides with the appearance of two peaks in the energy density. In this instance, the field $\chi$ scatters as a kink-antikink pair after being initially represented by a lump-like structure. We can also interpret the region A as composed of collision of two composite kinks that interpolate between -1 to 0 and 0 to 1 and two composite antikinks that interpolate between 1 to 0 and 0 to -1. As a result, we can see in some collisions the scattering of a kink-antikink pair interpolating between $0$ and $-1$ vacua whereas the pair interpolating between  $1$ and $0$ vacua form a bion state.

%%%%%%%%%%%%%%%%%%%%%%%%%%%%%%%%%%%%%%%%%%%%%%%%%%%%%%%%%%%%%%%%%%%%%
\begin{figure}
	\includegraphics[{angle=0,width=6cm}]{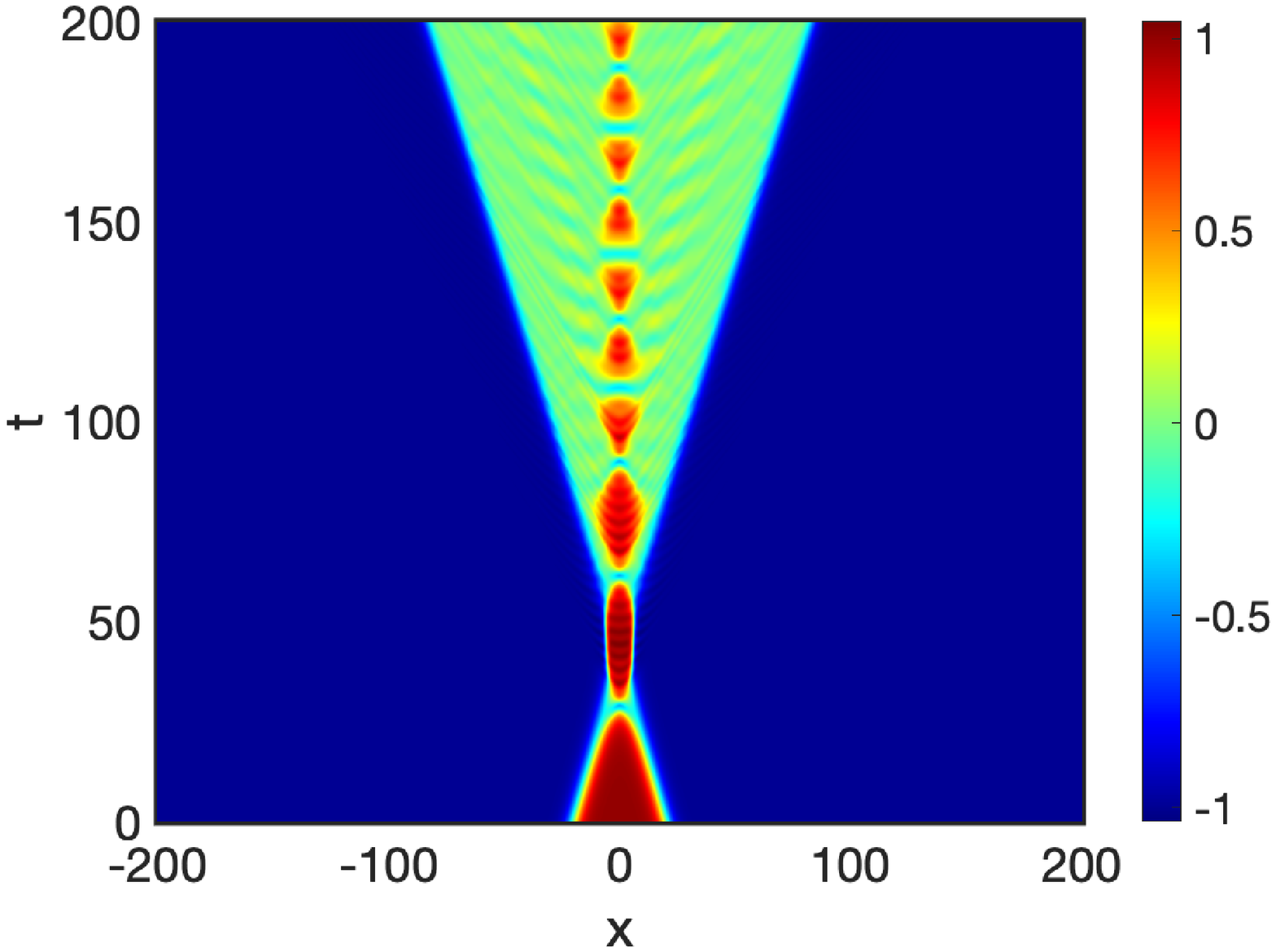}
	\includegraphics[{angle=0,width=6cm}]{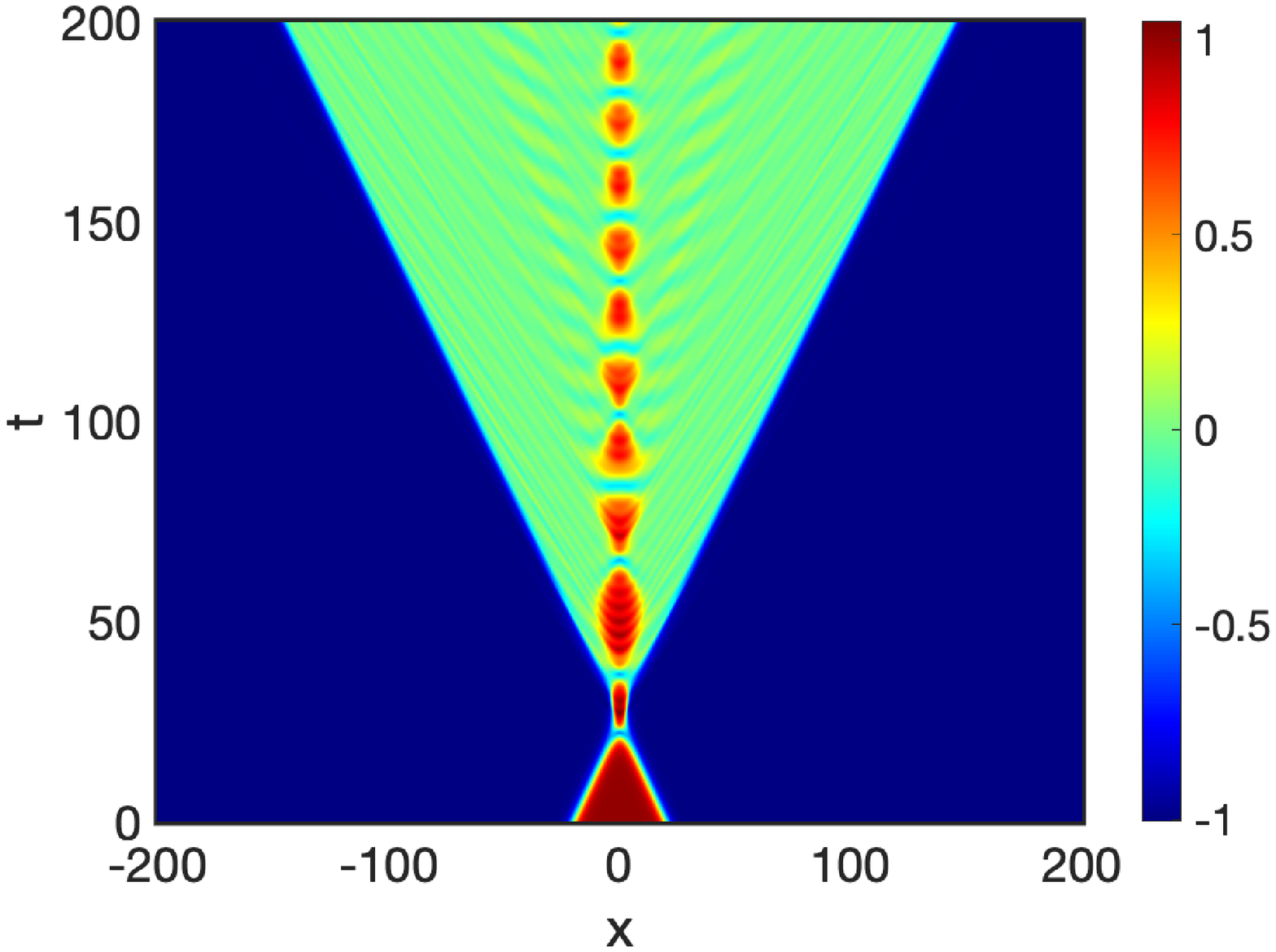}\\
	\includegraphics[{angle=0,width=6cm}]{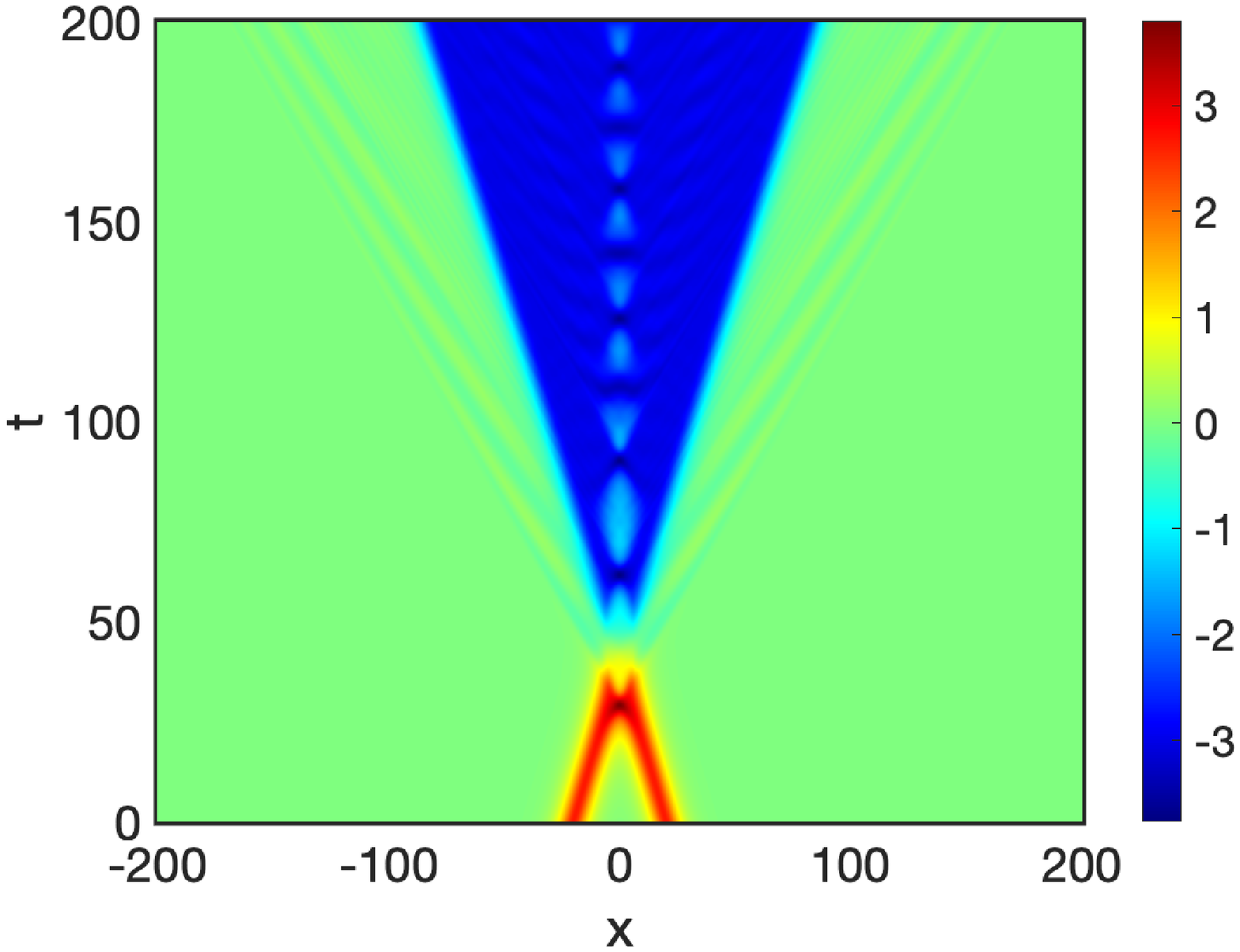}
	\includegraphics[{angle=0,width=6cm}]{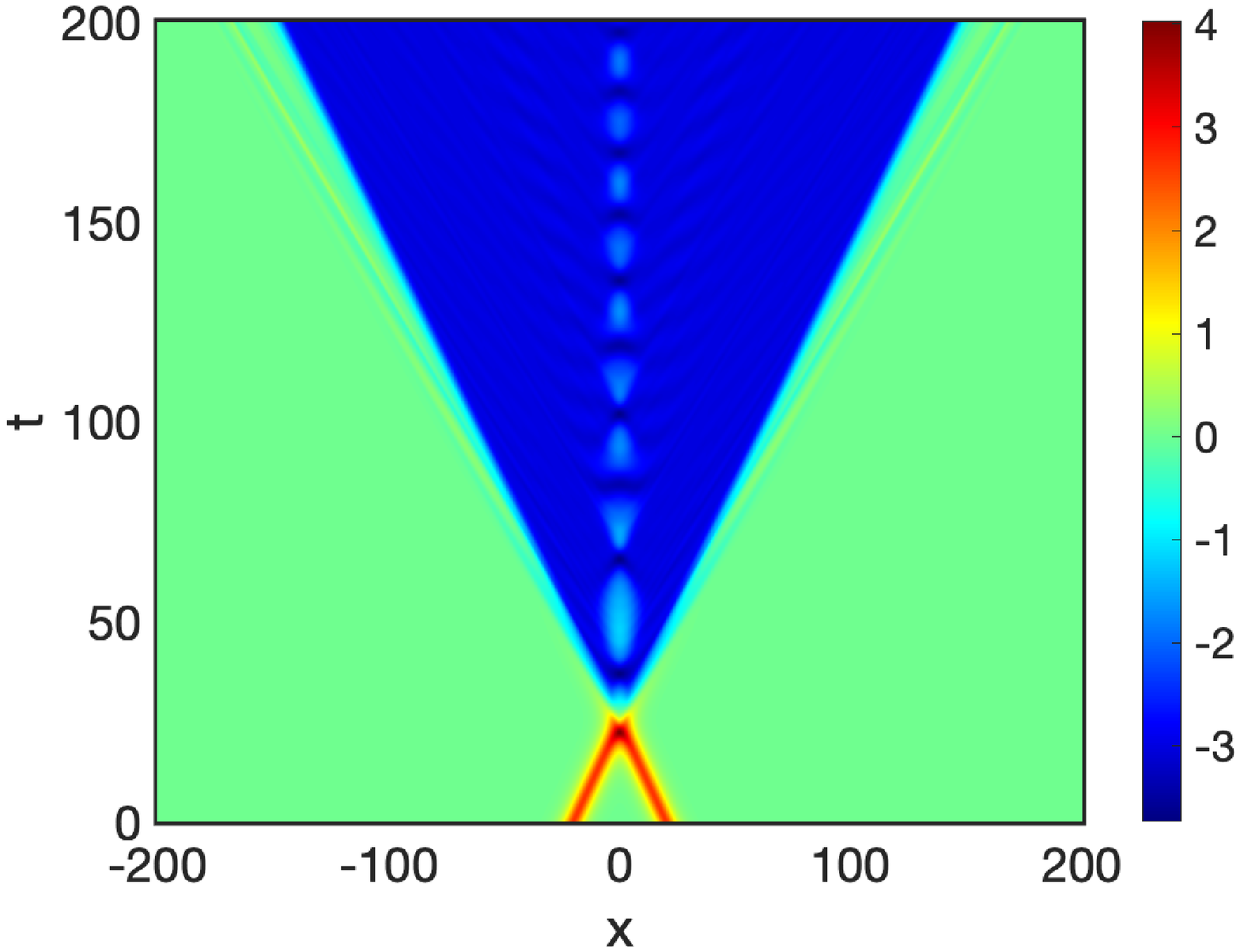}
	\caption{Region B -  $\phi$-component (top)  and $\chi$-component (bottom) for (a) $v=0.50$ with $r=0.11$ and (b) $v=0.80$ with $r=0.11$} 
\label {colB}
\end{figure}
%%%%%%%%%%%%%%%%%%%%%%%%%%%%%%%%%%%%%%%%%%%%%%%%%%%%%%%%%%%%%%%%%%%%%

We display the collisions for region B in Fig. \ref{colB}. For $x<0$ note that the $\phi$-component changes from $-1 \to 1$ to $-1 \to 0$, whereas the $\chi$-component changes to $0 \to 3 \to 0$ to $0 \to -3$, meaning that the $K_{21}$-kink changes to $K_{24}$ after scattering. For $x>0$ one can make a similar reasoning: the $\phi$-component changes from $1 \to -1$ to $ 0 \to -1$, and the $\chi$-component from  $0 \to 3 \to 0$  to $ -3 \to 0$, meaning that the $\bar K_{21}$-antikink changes to $\bar K_{24}$ after scattering. Then, the collision can be characterized by $K_{21}+\bar K_{21} \to K_{24} + \bar K_{24}$ and we noted that the emission of radiation is more evidently produced by the $\chi$-component. The scattering of the $\phi$-component in both regions A and B reveals the formation of a kink-antikink pair, connecting minima 0 and 1 as well as the production of oscillating pulses around $x=0$. In contrast to region A, increasing the initial velocity causes a change in the collision outcome of the $\chi$-component. After the collision, the $\chi$-component in region B shows the formation of an antikink-kink pair.

%%%%%%%%%%%%%%%%%%%%%%%%%%%%%%%%%%%%%%%%%%%%%%%%%%%%%%%%%%%%%%%%%%%%%
\begin{figure}
	\includegraphics[{angle=0,width=5cm}]{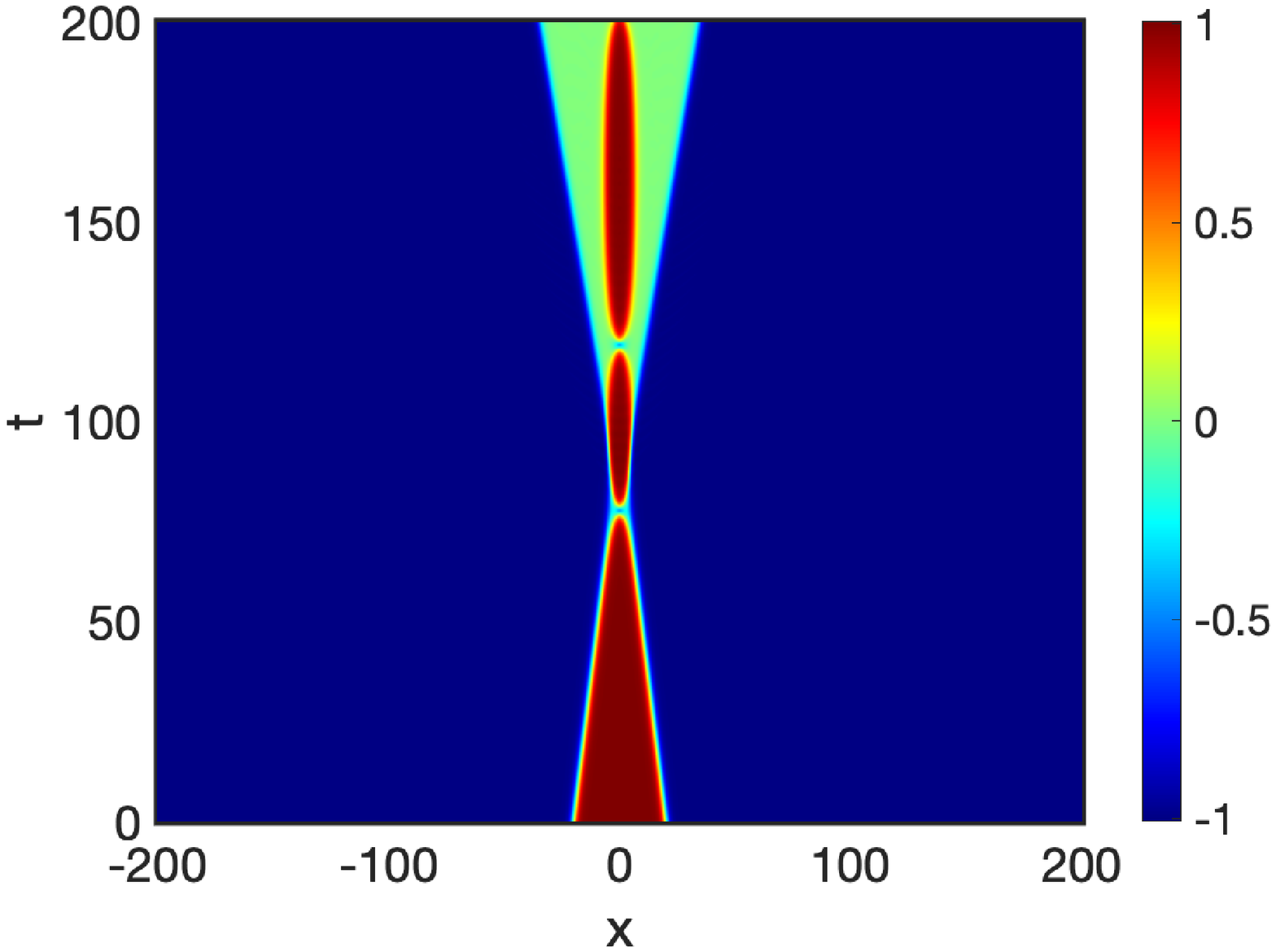}
	\includegraphics[{angle=0,width=5cm}]{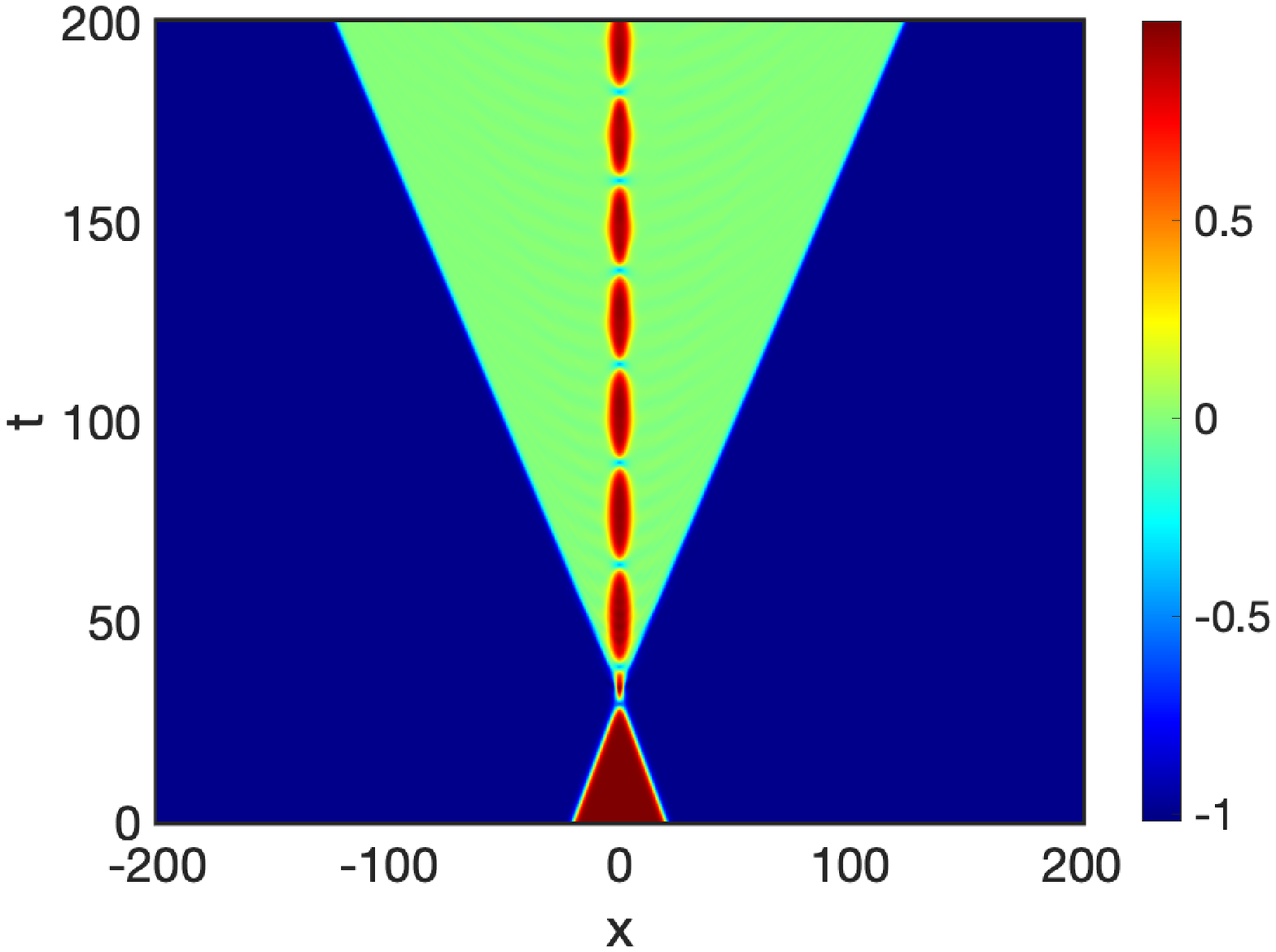}
	\includegraphics[{angle=0,width=5cm}]{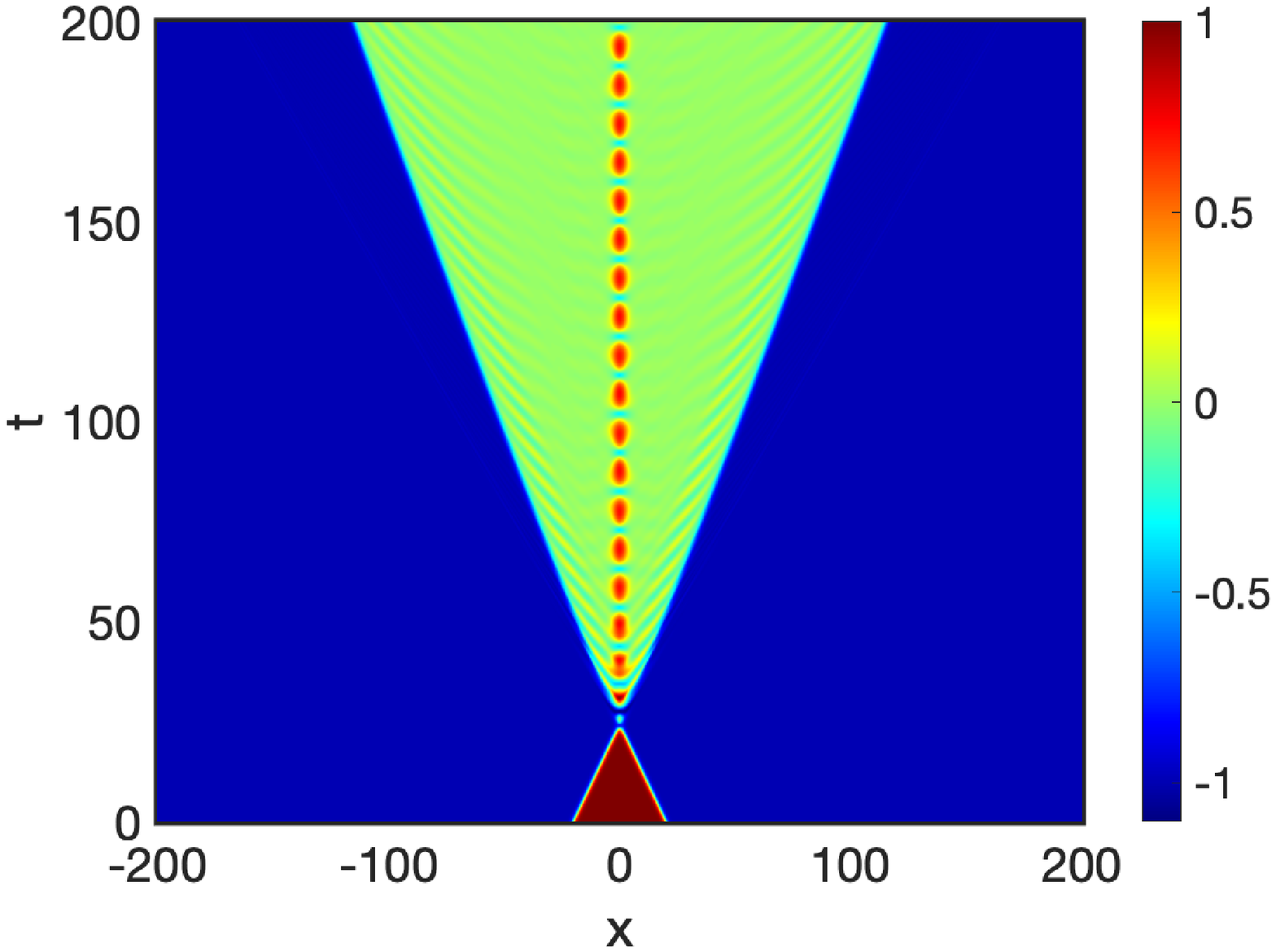}\\
\includegraphics[{angle=0,width=5cm}]{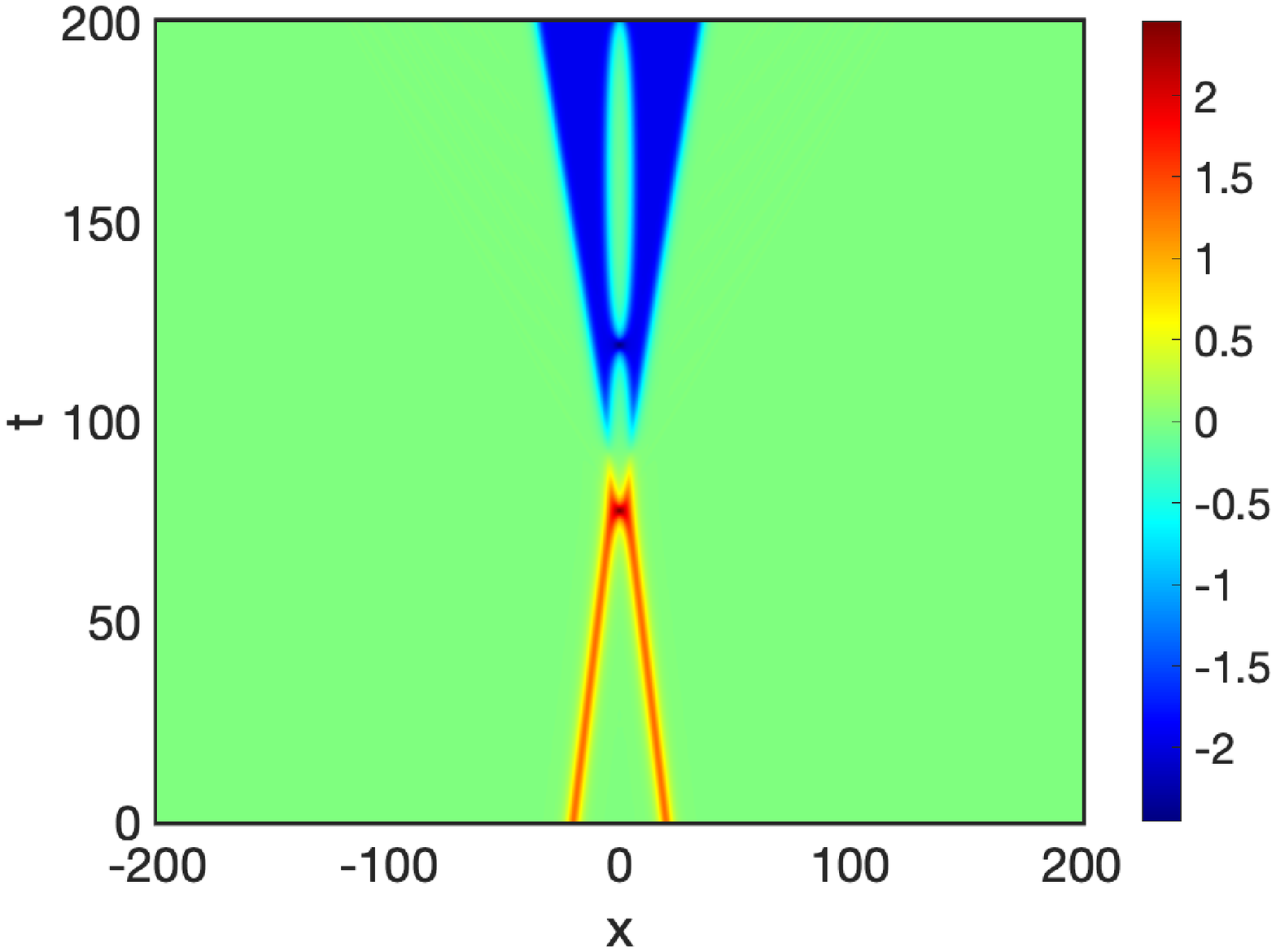}
	\includegraphics[{angle=0,width=5cm}]{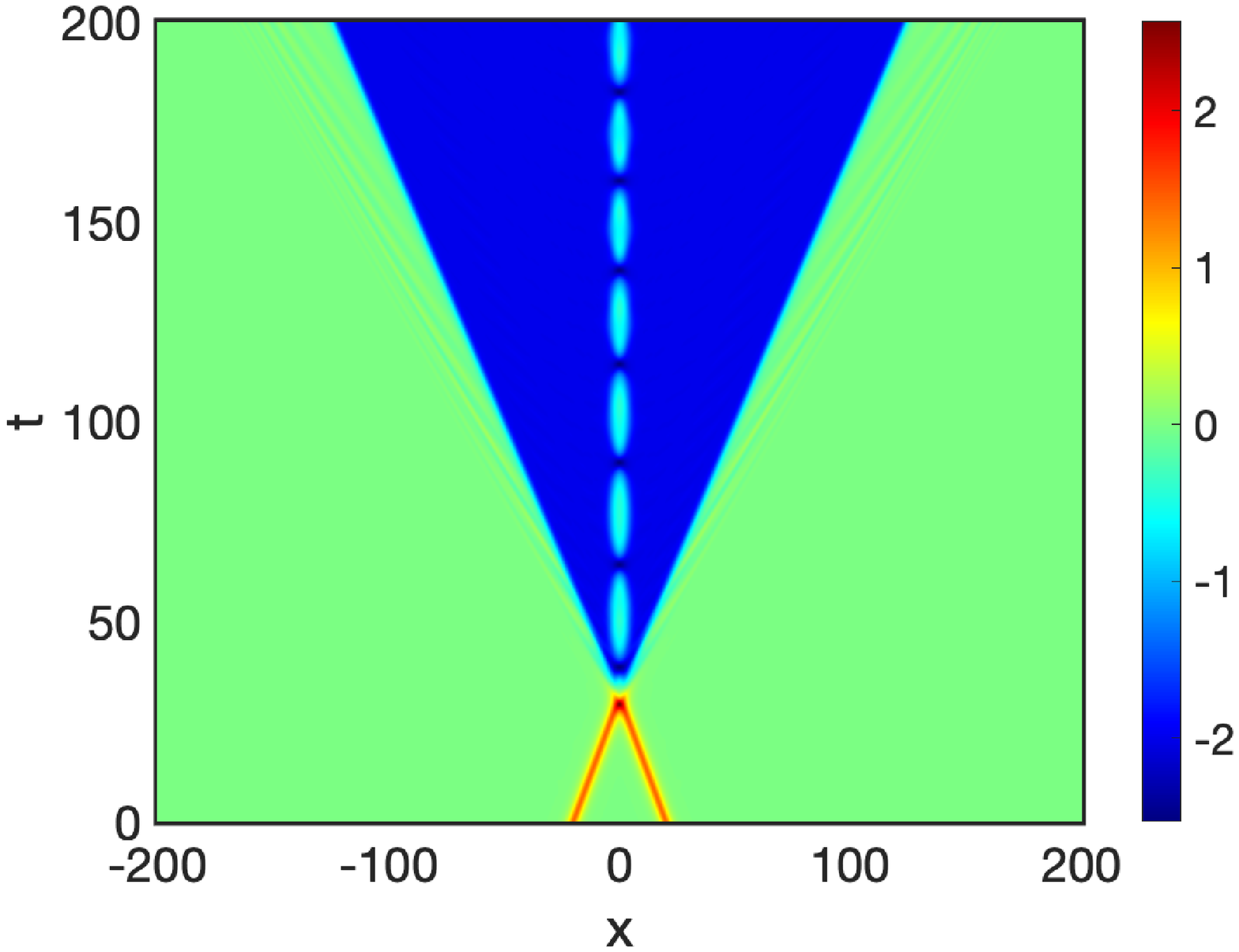}
	\includegraphics[{angle=0,width=5cm}]{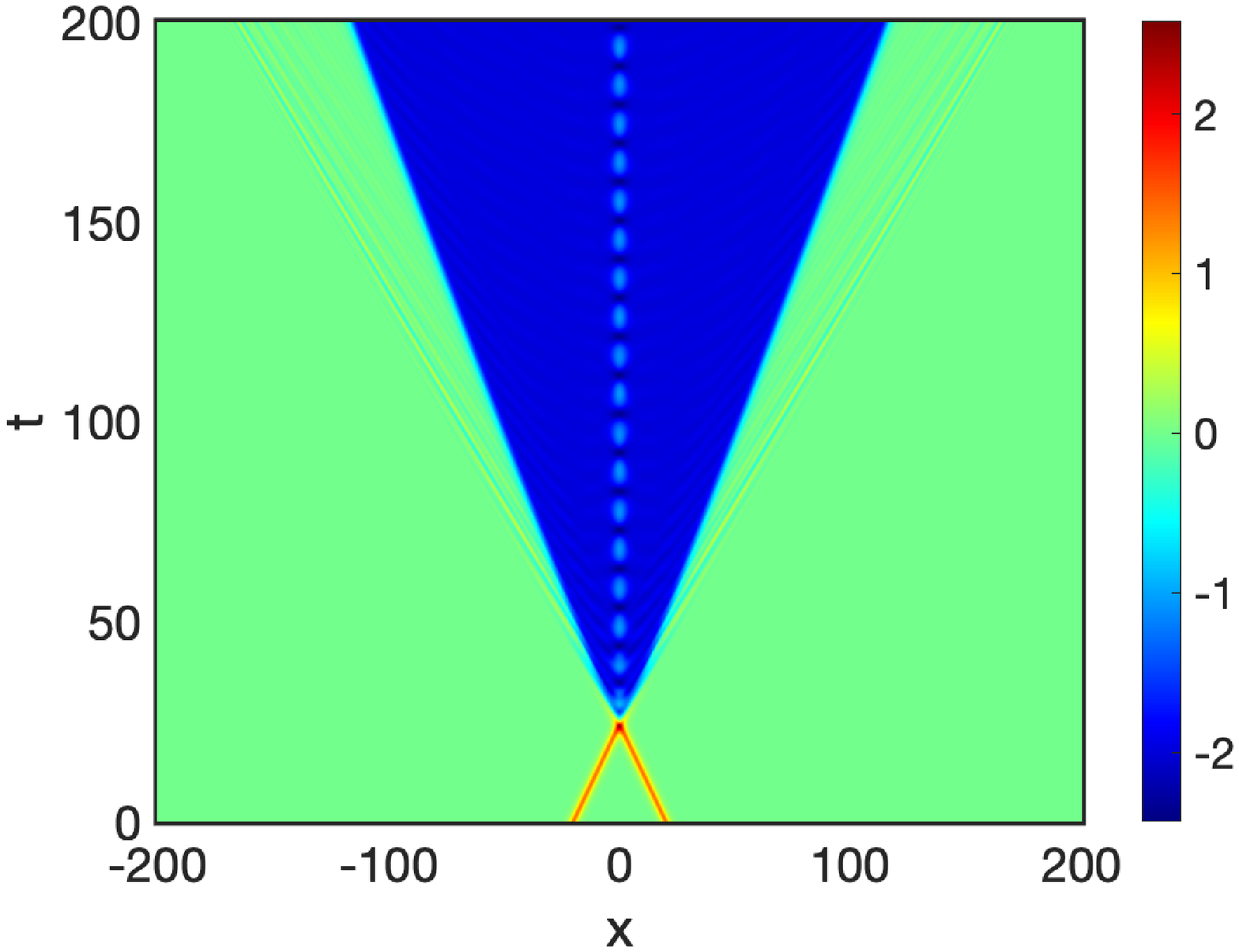}
	\caption{Region C - $\phi$-component (top)  and $\chi$-component (bottom) for  (a) $v=0.21$ with $r=0.27$, (b) $v=0.64$ with $r=0.25$ and (c) $v=0.82$ with $r=0.255$. }
\label {colC}
\end{figure}
%%%%%%%%%%%%%%%%%%%%%%%%%%%%%%%%%%%%%%%%%%%%%%%%%%%%%%%%%%%%%%%%%%%%%

The Figs. \ref{colC} depicts the collision of the region C. The results from this region are very similar from those in region B, since they are also characterized by $K_{21}+\bar K_{21} \to K_{24} + \bar K_{24}$. The $\chi$-component oscillates around $x=0$, which is the main difference. We notice that the oscillating pulses reach the vacuum $\phi=0$ in region C. This behavior is not observed in region B, where the central oscillations revolve around the vacuum $\phi=-1$. Region C is significant because it marks the start of the shift in the energy density behavior from two peaks to one peak centered at $x=0$.

%%%%%%%%%%%%%%%%%%%%%%%%%%%%%%%%%%%%%%%%%%%%%%%%%%%%%%%%%%%%%%%%%%%%%
\begin{figure}
	\includegraphics[{angle=0,width=6cm}]{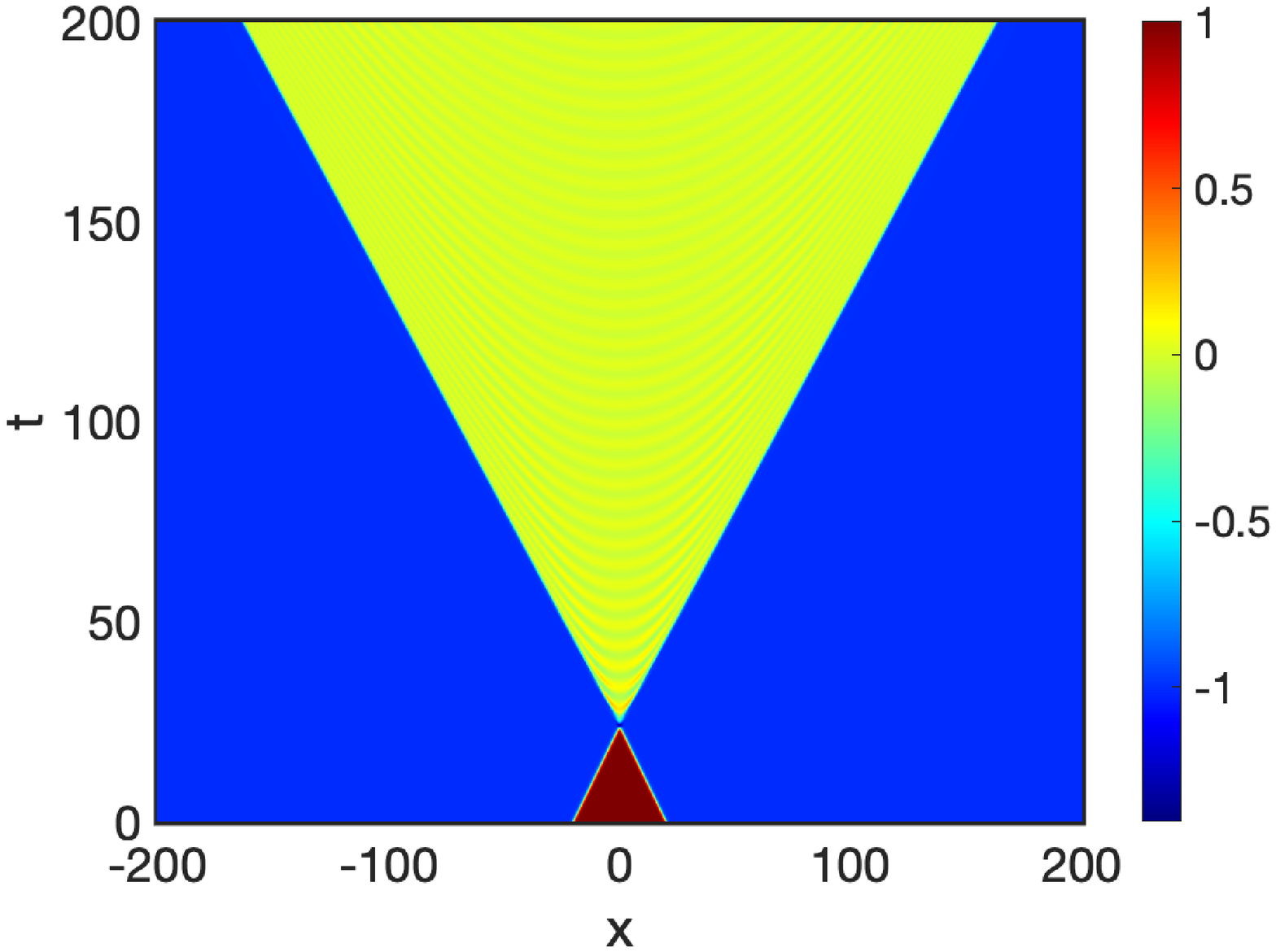}
	\includegraphics[{angle=0,width=6cm}]{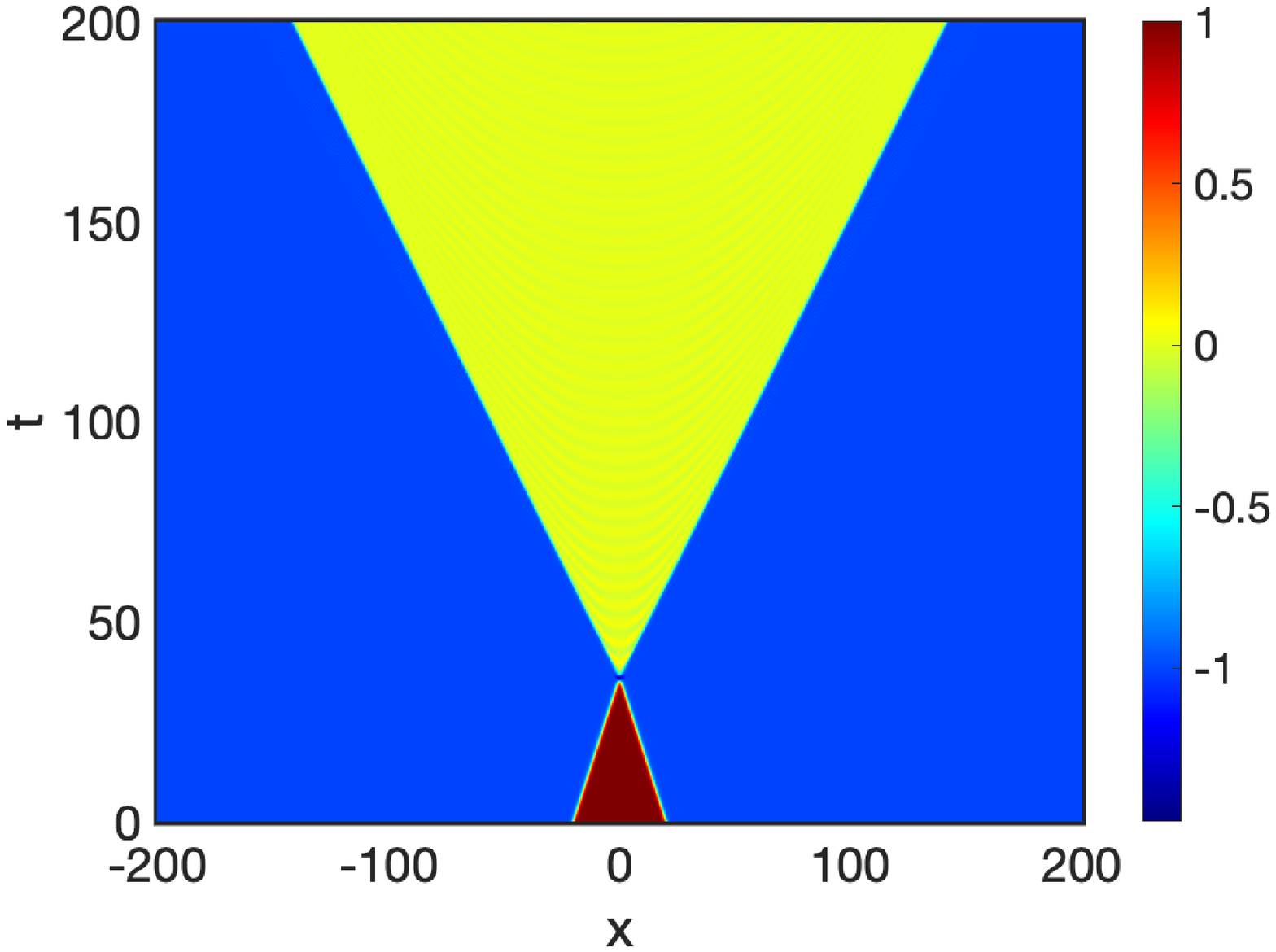} \\
	\includegraphics[{angle=0,width=6cm}]{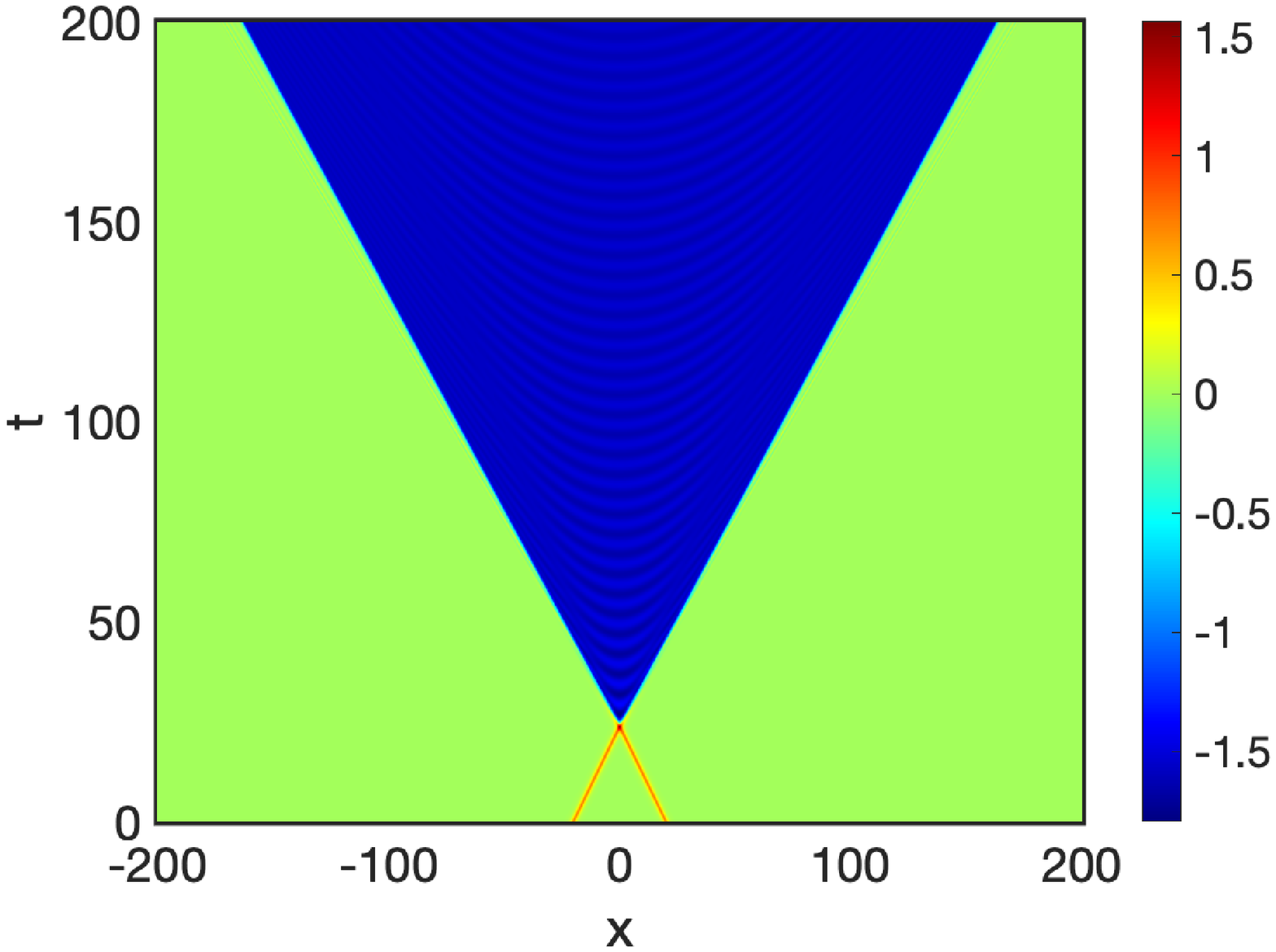}
	\includegraphics[{angle=0,width=6cm}]{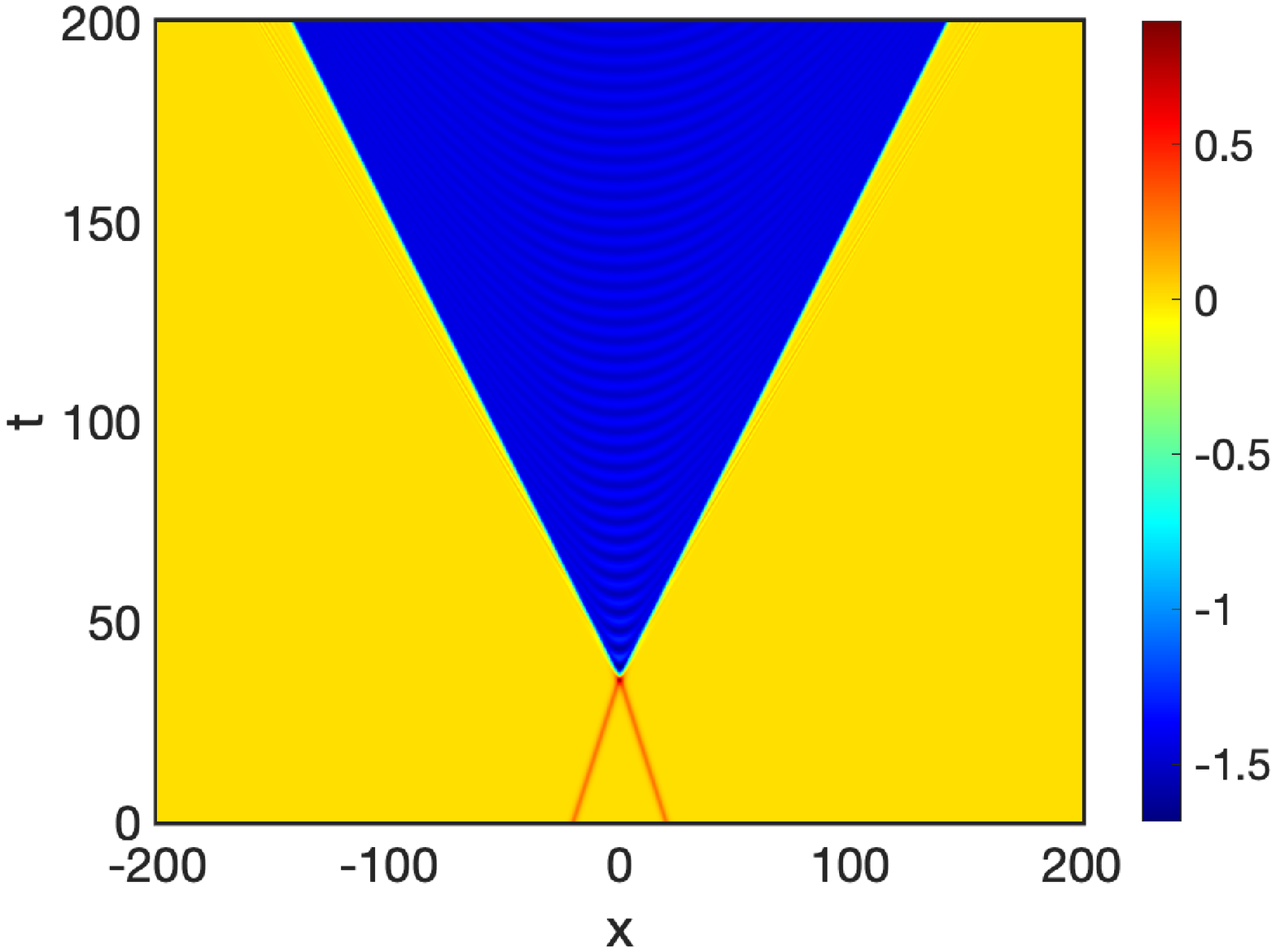}
	\caption{Region D - $\phi$-component (top)  and $\chi$-component (bottom) for (a) $v=0.84$ with $r=0.40$ and (b) $v=0.555$ with $r=0.483$. }
\label{colD}
\end{figure}

%%%%%%%%%%%%%%%%%%%%%%%%%%%%%%%%%%%%%%%%%%%%%%%%%%%%%%%%%%%%%%%%%%%%%

The region D has collisions represented by the Figs. \ref{colD}. The results from this region are very similar from those in region B, since they are also characterized by $K_{21}+\bar K_{21} \to K_{24} + \bar K_{24}$. Oscillations around $x=0$, on the other hand, are not observed in either the $\phi$ or $\chi$ scattering components. This region corresponds to a range for larger $v$ and $r$ values. As a result, the internal structure contributes less to the collision process, yielding simpler results, particularly with only the phase change after the collision.

%%%%%%%%%%%%%%%%%%%%%%%%%%%%%%%%%%%%%%%%%%%%%%%%%%%%%%%%%%%%%%%%%%%%%
\begin{figure}
	\includegraphics[{angle=0,width=6cm}]{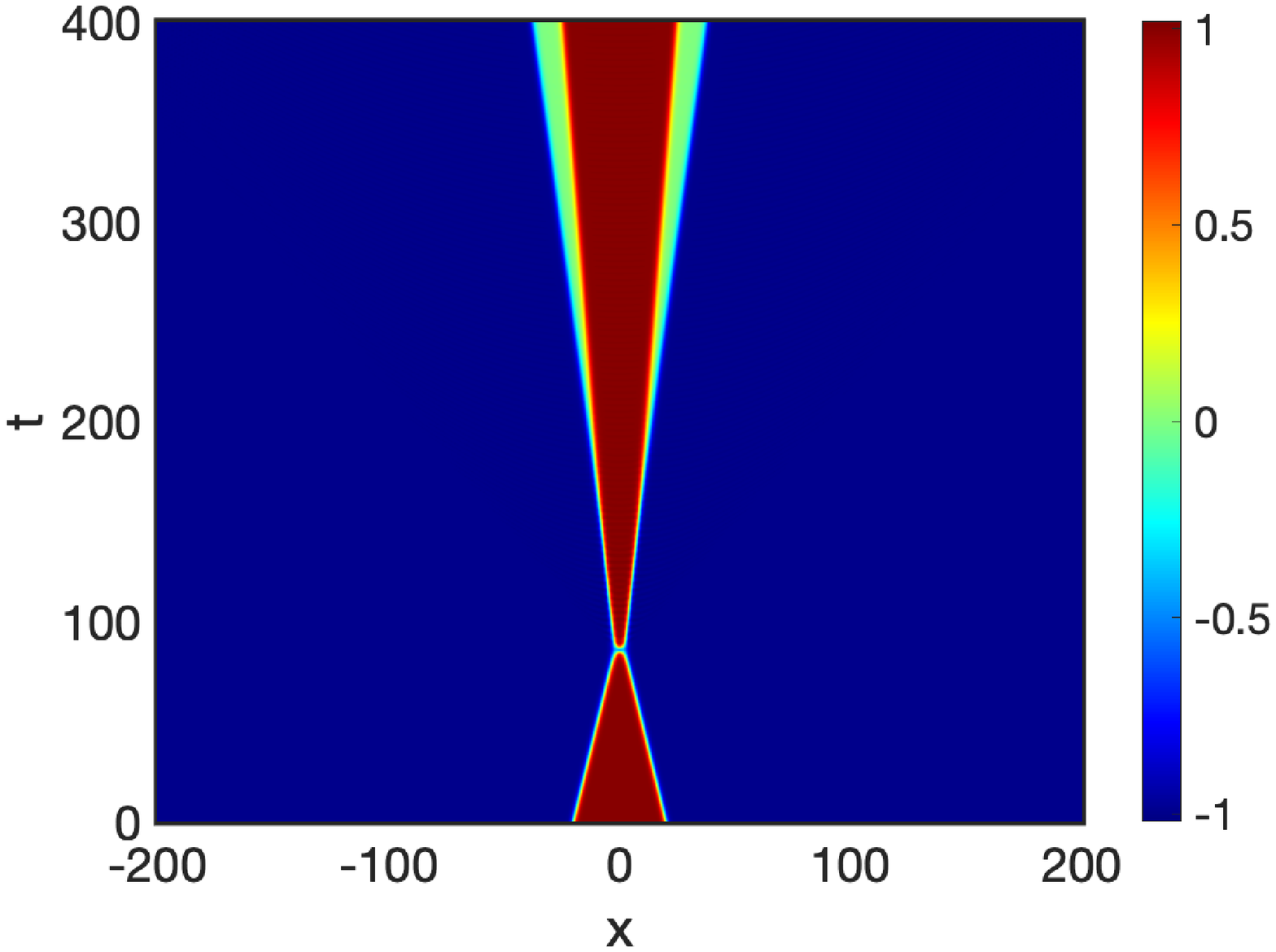}
	\includegraphics[{angle=0,width=6cm}]{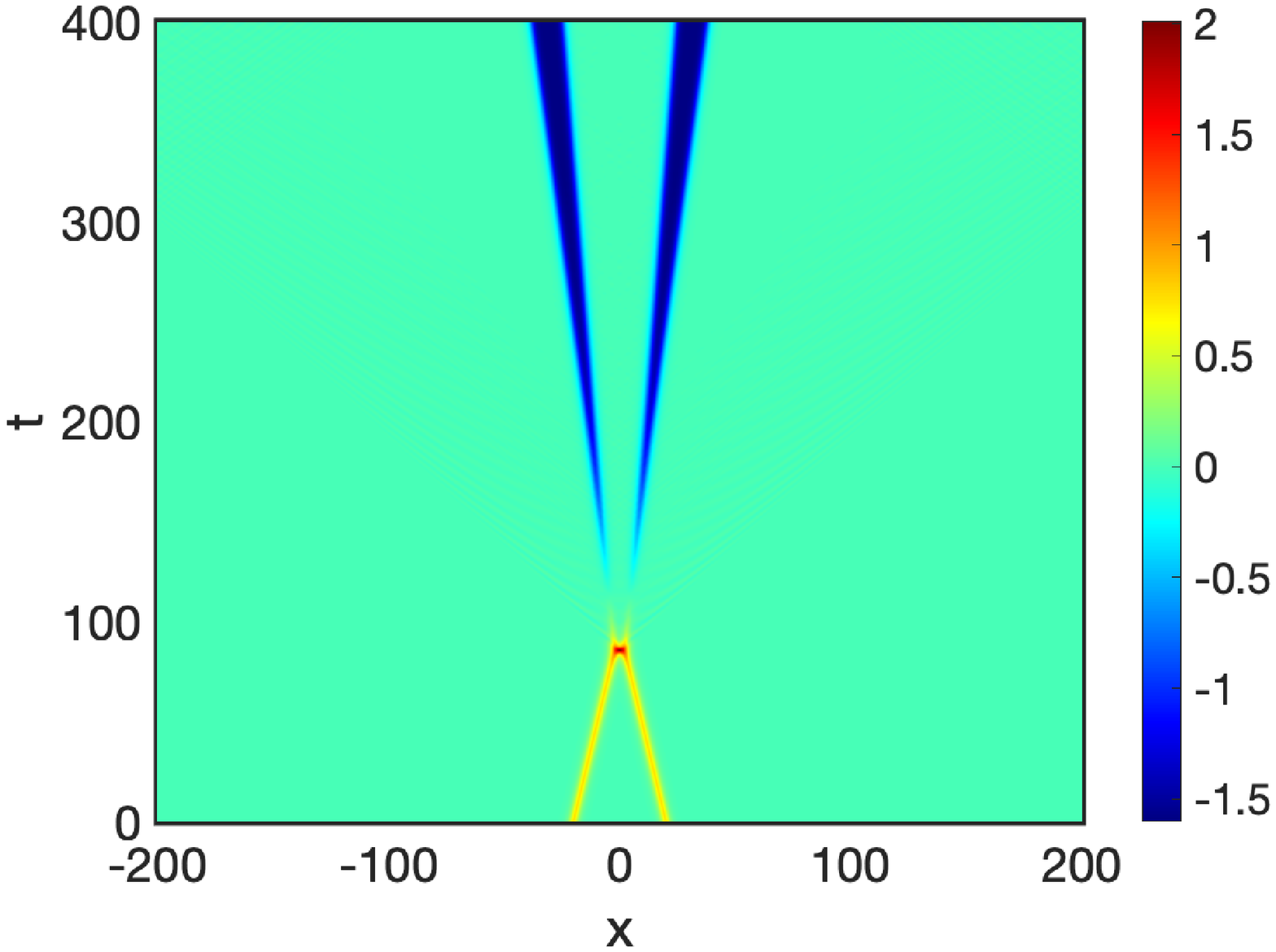}
	\caption{Region E - $\phi$-component (left) and $\chi$-component (right) for $v=0.208$ with $r=0.392$. }
\label{colE}
\end{figure}
%%%%%%%%%%%%%%%%%%%%%%%%%%%%%%%%%%%%%%%%%%%%%%%%%%%%%%%%%%%%%%%%%%%%%

For values $0.3<r<0.5$ and as the initial velocity decreases, we observe the formation of a complex structure that corresponds to regions E, F, G, H, and I. An intriguing illustration of the region E scattering can be found in Fig. \ref{colE}. Note that the $\chi$-component changes from $0 \to 0$ to $0 \to -1 \to 0$. According to this finding, two antikink-kink pairs form in the lump-lump collision of the $\chi$-component. As opposed to that, the $\phi$-component changes from $-1 \to 1$ to $-1 \to 0 \to 1$ for $x<0$, demonstrating that the kink-antikink collision promotes the appearance of a double kink. 

\begin{figure}
	\includegraphics[{angle=0,width=6cm}]{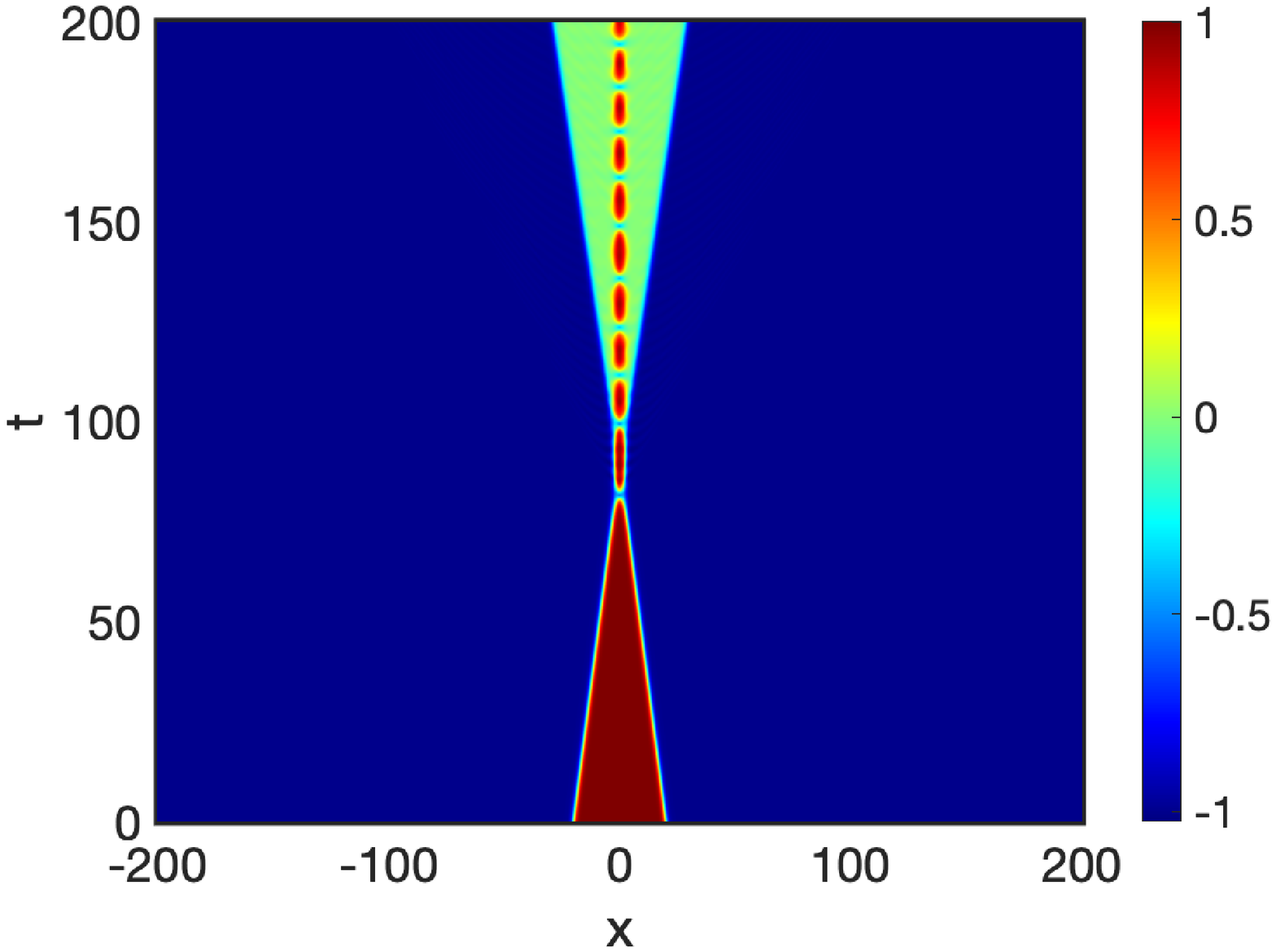}
	\includegraphics[{angle=0,width=6cm}]{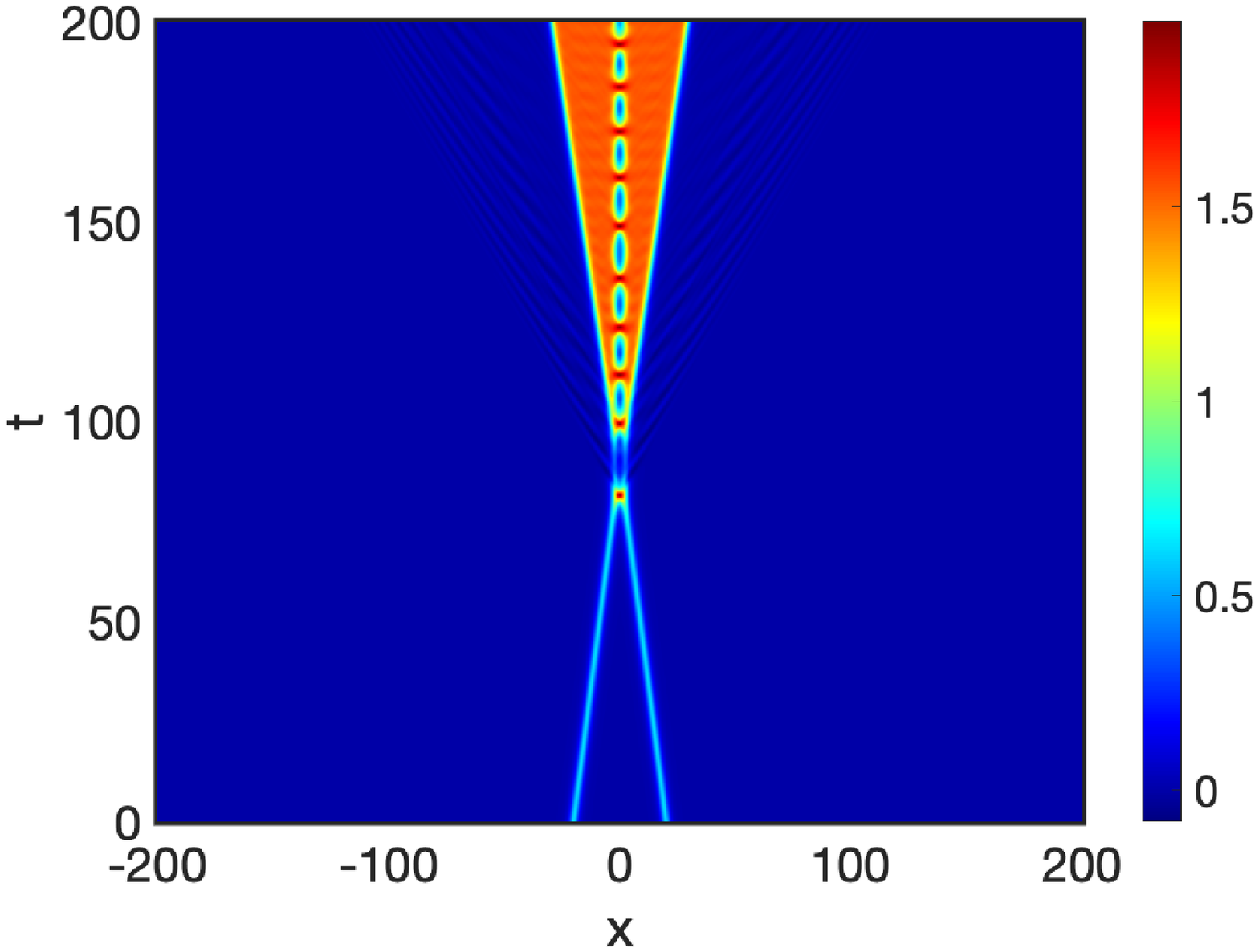}
	\caption{Region F - $\phi$-component (left) and $\chi$-component (right) for $v=0.223$ with $r=0.42$. }
\label{colF}
\end{figure}

We notice the formation of the region F at low velocities but with a small increase in the parameter $r$. This region has collisions represented by the Figs. \ref{colF}. This region is characterized by scattering of the type $K_{21}+\bar K_{21} \to K_{23} + \bar K_{23}$ and a central oscillations around $x=0$. Note that, contrary to what observed in the region A, the oscillations are very localized, with no significant distortion.

%%%%%%%%%%%%%%%%%%%%%%%%%%%%%%%%%%%%%%%%%%%%%%%%%%%%%%%%%%%%%%%%%%%%%
%%%%%%%%%%%%%%%%%%%%%%%%%%%%%%%%%%%%%%%%%%%%%%%%%%%%%%%%%%%%%%%%%%%%%

\begin{figure}
	\includegraphics[{angle=0,width=6cm}]{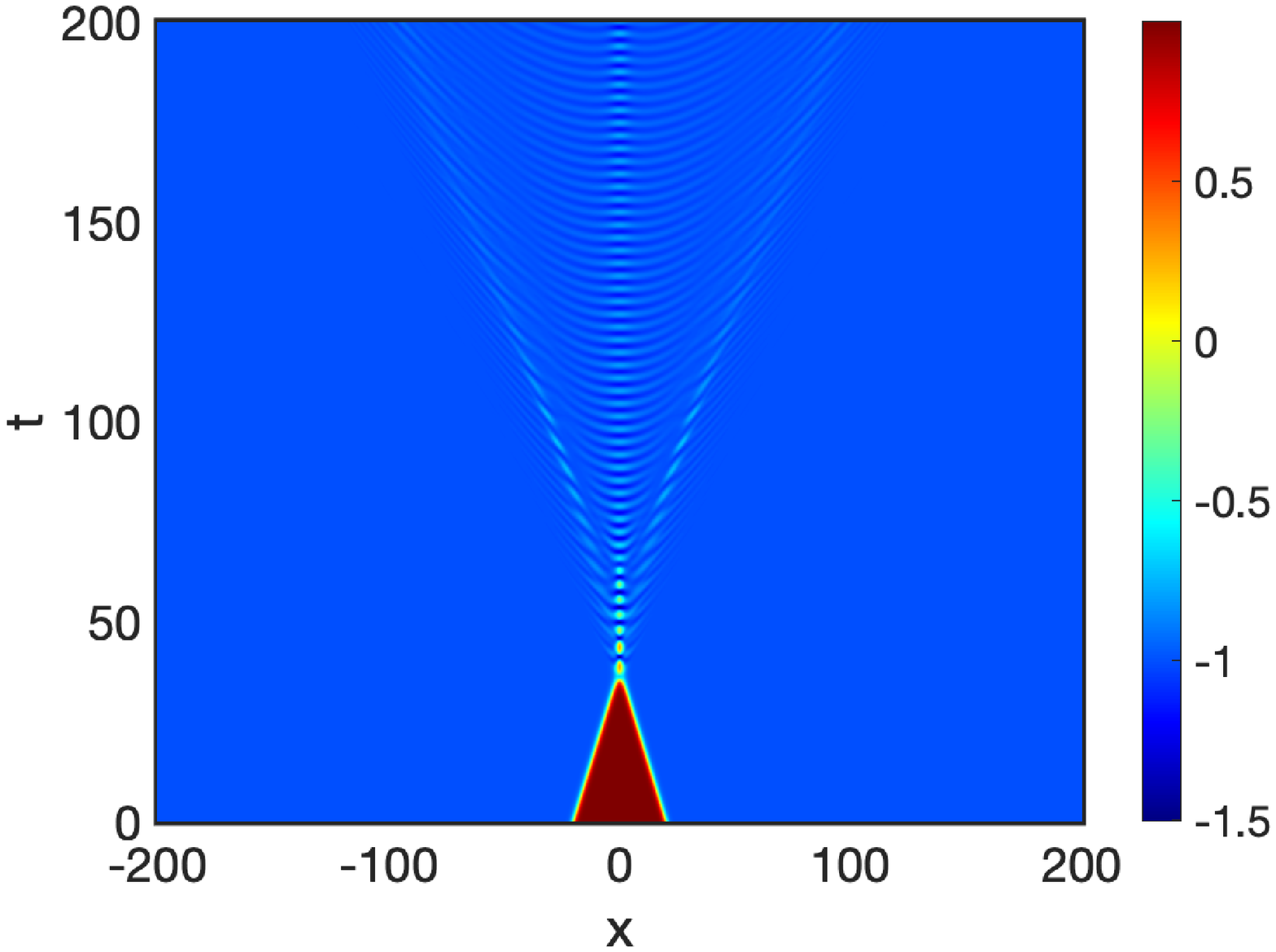}
	\includegraphics[{angle=0,width=6cm}]{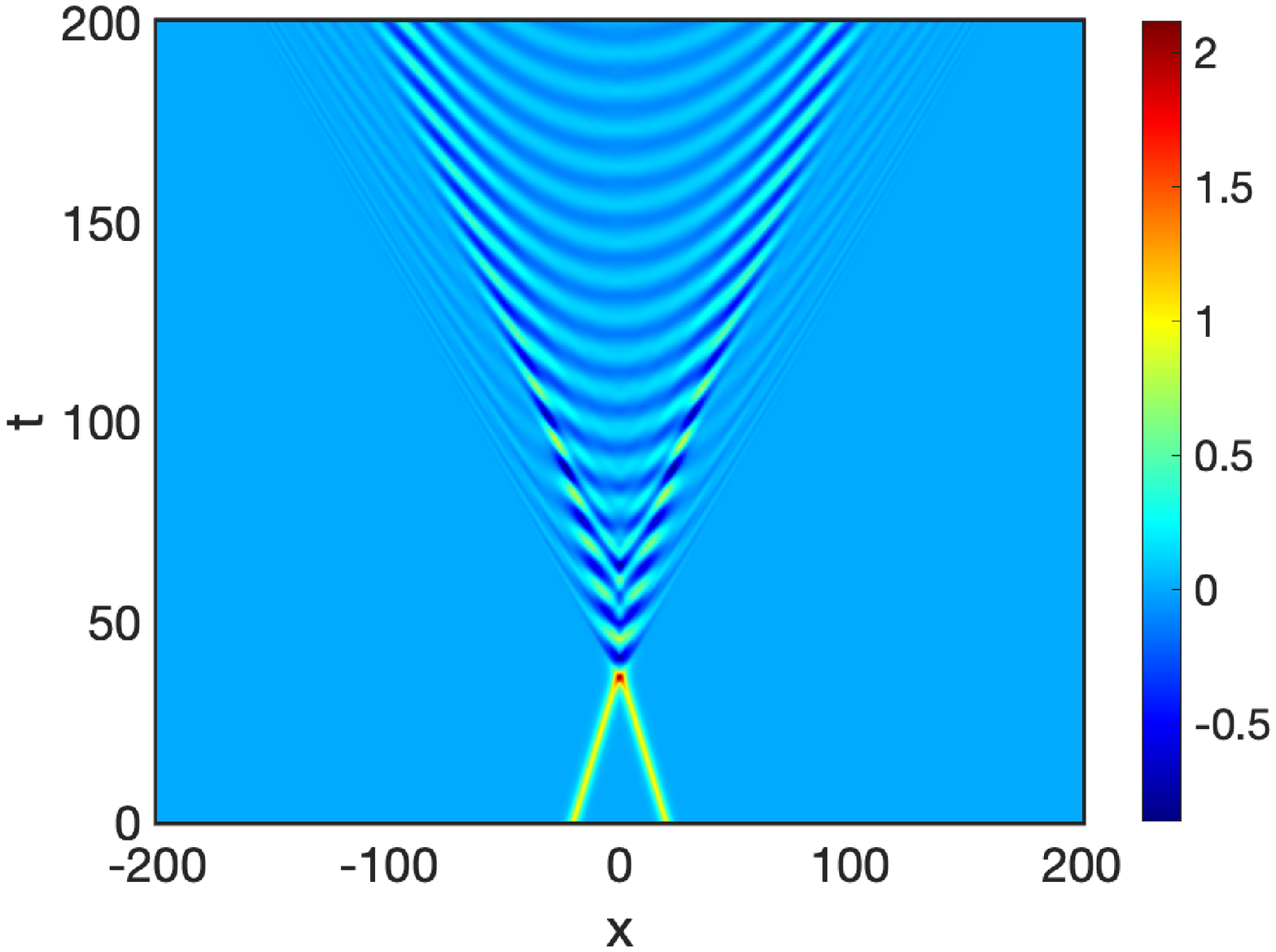}
	\caption{Region G - $\phi$-component (left) and $\chi$-component (right) for $v=0.524$ with $r=0.336$. }
\label{colG}
\end{figure}

\begin{figure}
	\includegraphics[{angle=0,width=6cm}]{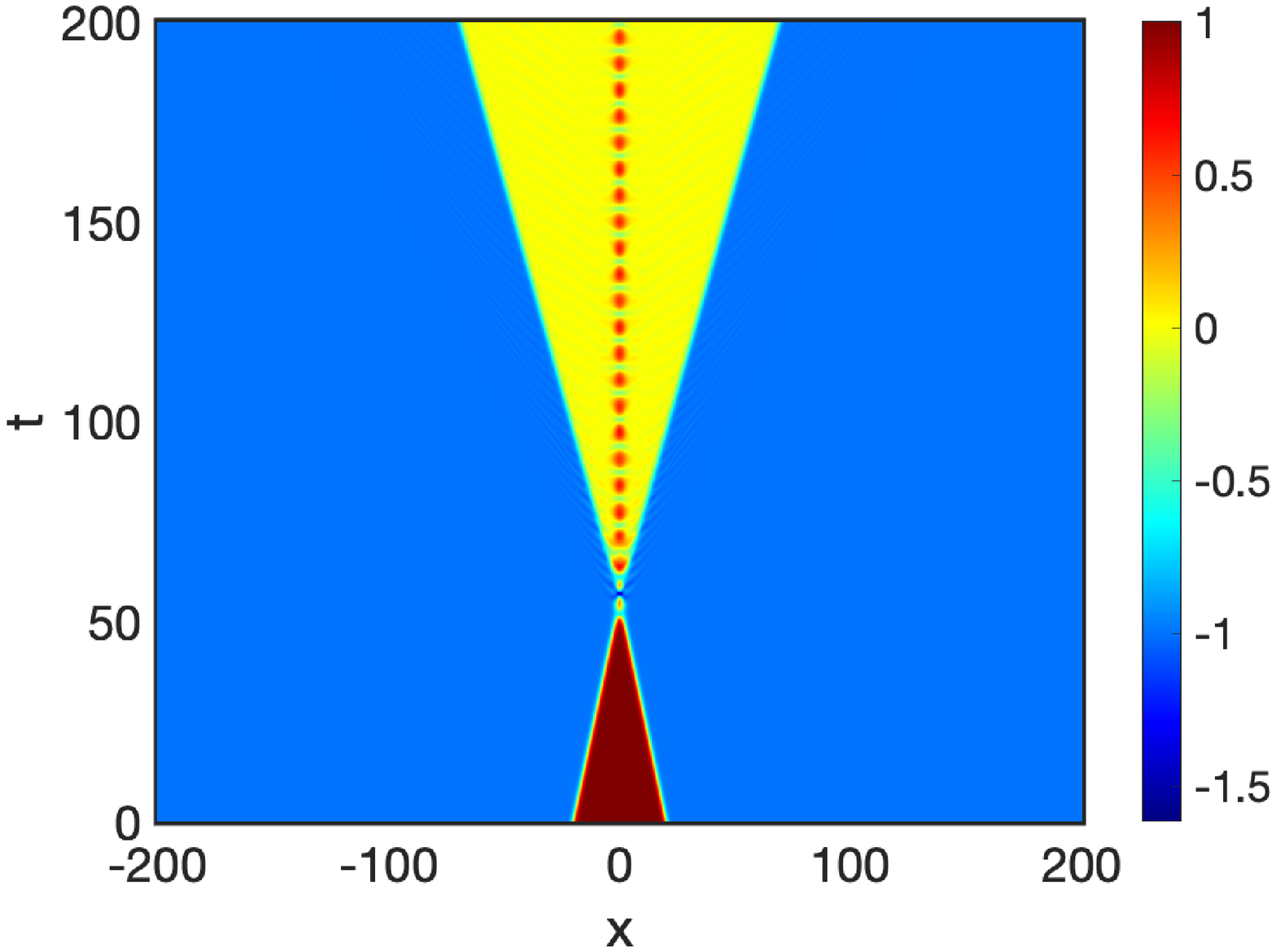}
	\includegraphics[{angle=0,width=6cm}]{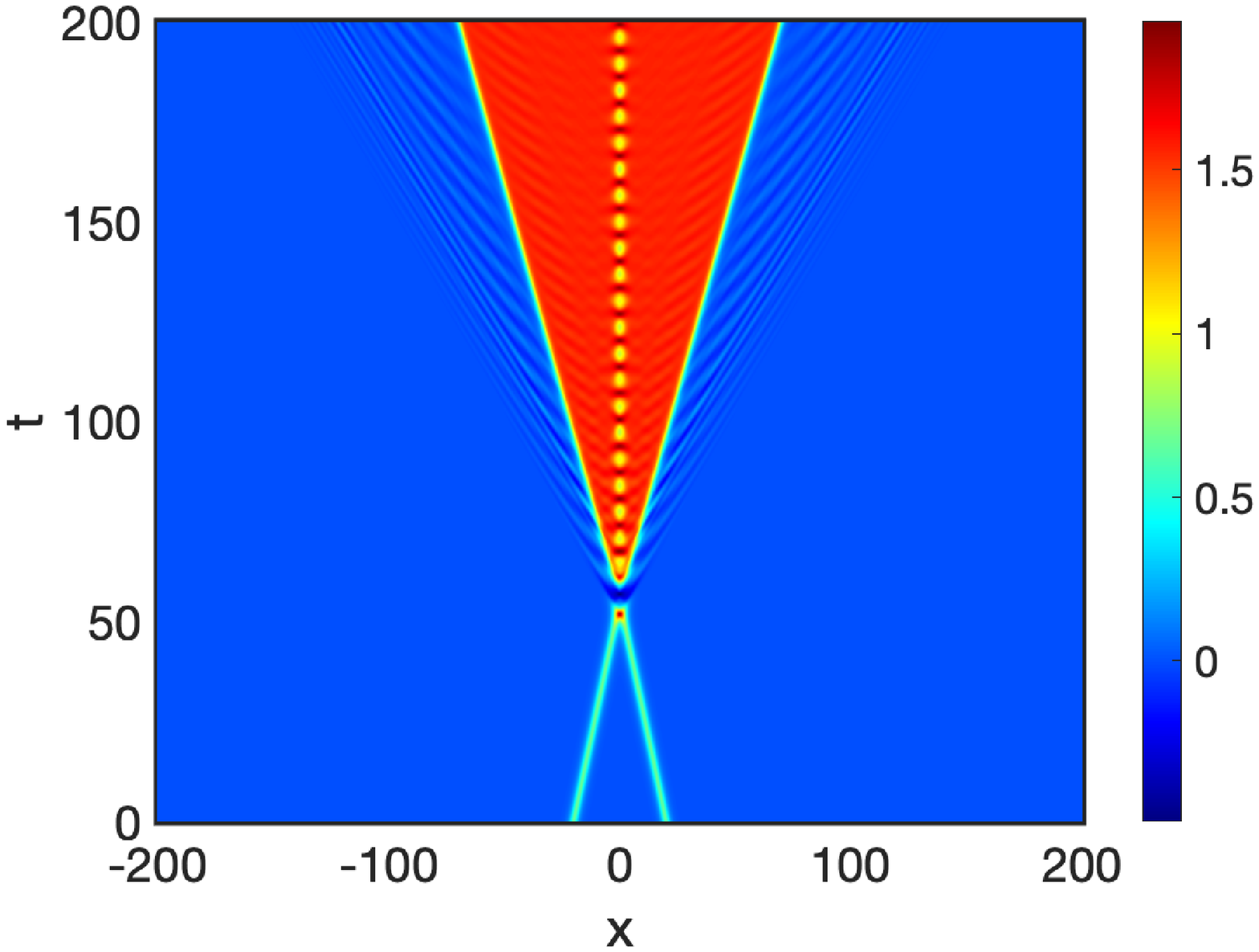}
	\caption{Region H - $\phi$-component (left) and $\chi$-component (right) for (a) $v=0.3618$ with $r=0.4085$. }
\label{colH}
\end{figure}

\begin{figure}
	\includegraphics[{angle=0,width=6cm}]{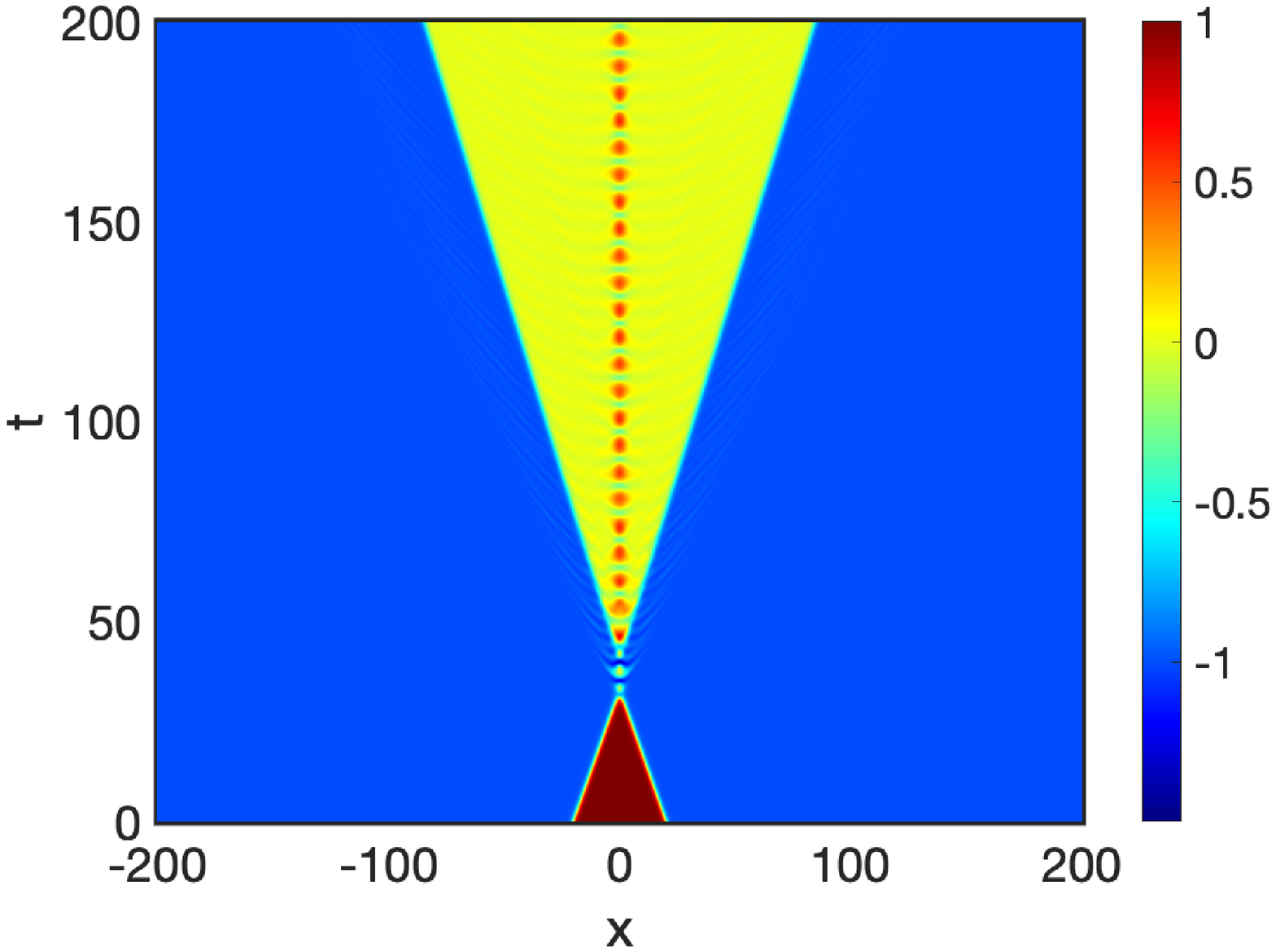}
	\includegraphics[{angle=0,width=6cm}]{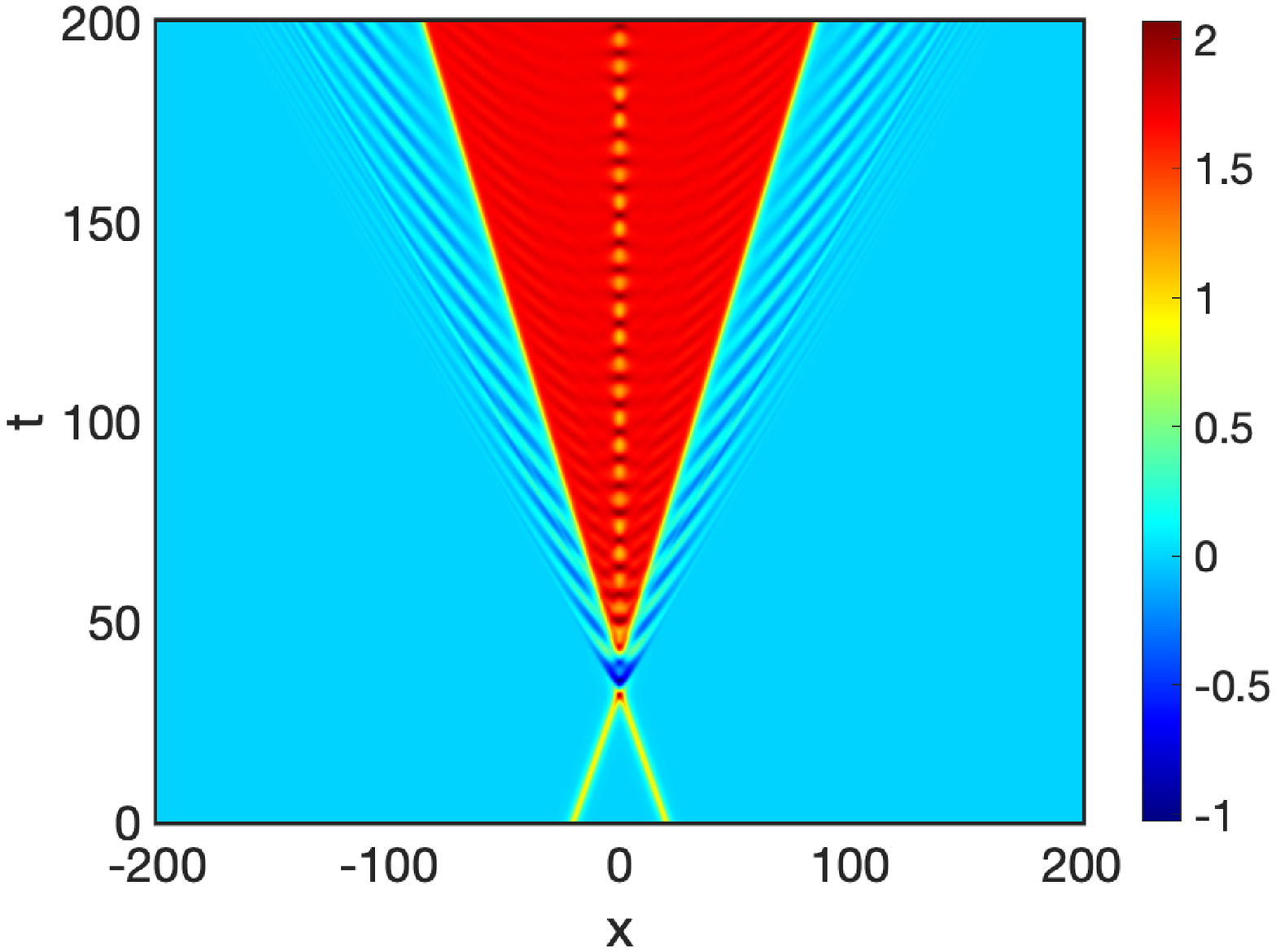}
	\caption{Region I - $\phi$-component (left) and $\chi$-component (right) for $v=0.0.609$ with $r=0.346$. }
\label{colI}
\end{figure}

The collisions for region G are plotted in Fig. \ref{colG}. In this region the defects annihilate, with the fields restoring to the vacuum $(\phi,\chi)=(-1,0) $. The $\phi$-component produces two symmetric radiation jets and a localized oscillation around $x=0$. The $\chi$ field produces delocalized radiation. The H (Figs. \ref{colH}) and I (Figs. \ref{colI}) regions behave similarly to that of the F region.

%%%%%%%%%%%%%%%%%%%%%%%%%%%%%%%%%%%%%%%%%%%%%%%%%%%%%%%%%%%%%%%%%%%%%
%%%%%%%%%%%%%%%%%%%%%%%%%%%%%%%%%%%%%%%%%%%%%%%%%%%%%%%%%%%%%%%%%%%%%

%%%%%%%%%%%%%%%%%%%%%%%%%%%%%%%%%%%%%%%%%%%%%%%%%%%%%%%%%%%%%%%%%%%%%

%%%%%%%%%%%%%%%%%%%%%%%%%%%%%%%%%%%%%%%%%%%%%%%%%%%%%%%%%%%%%%%%%%%%%

\begin{figure}
\includegraphics[{angle=0,width=6cm}]{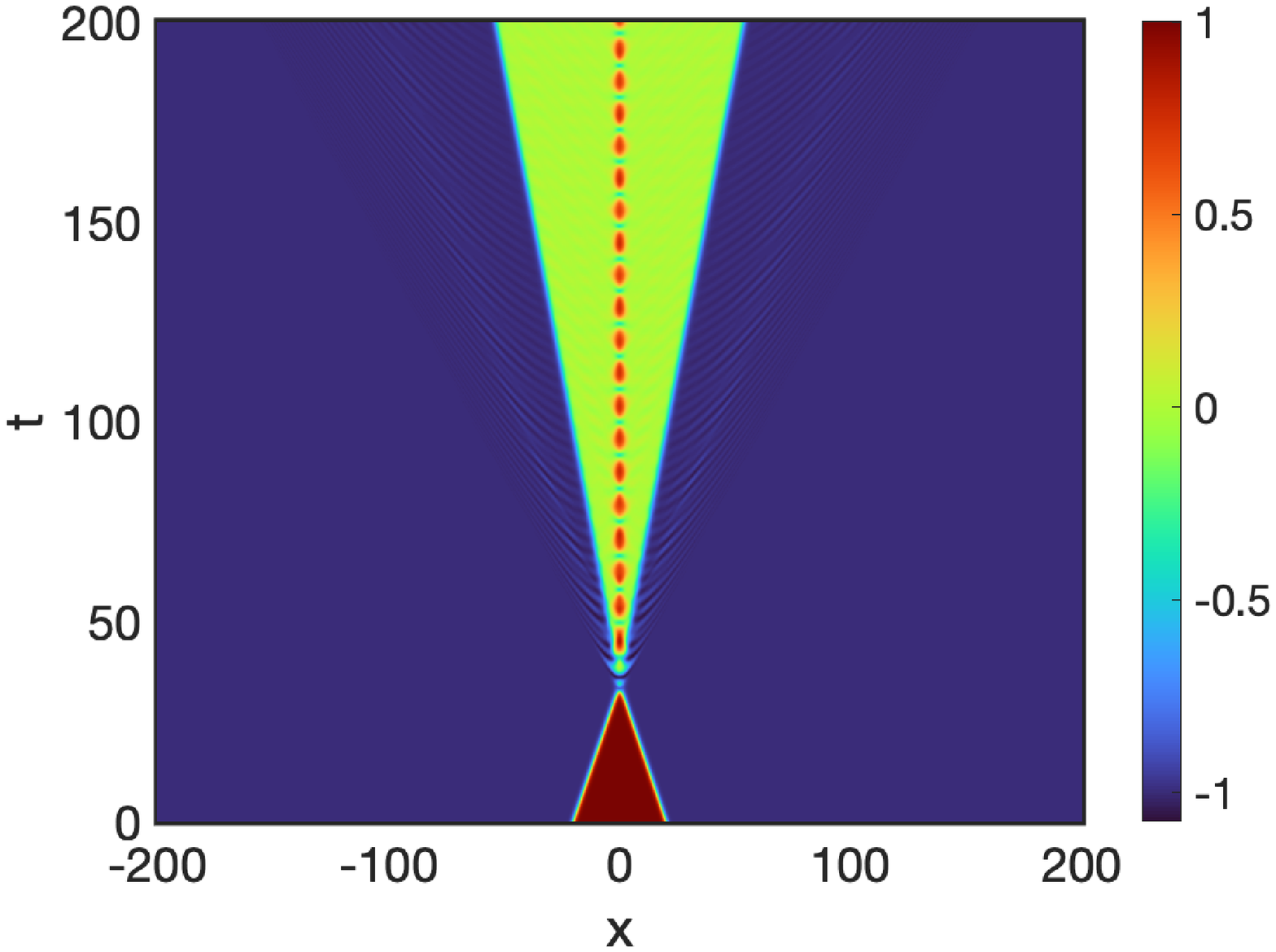}
\includegraphics[{angle=0,width=6cm}]{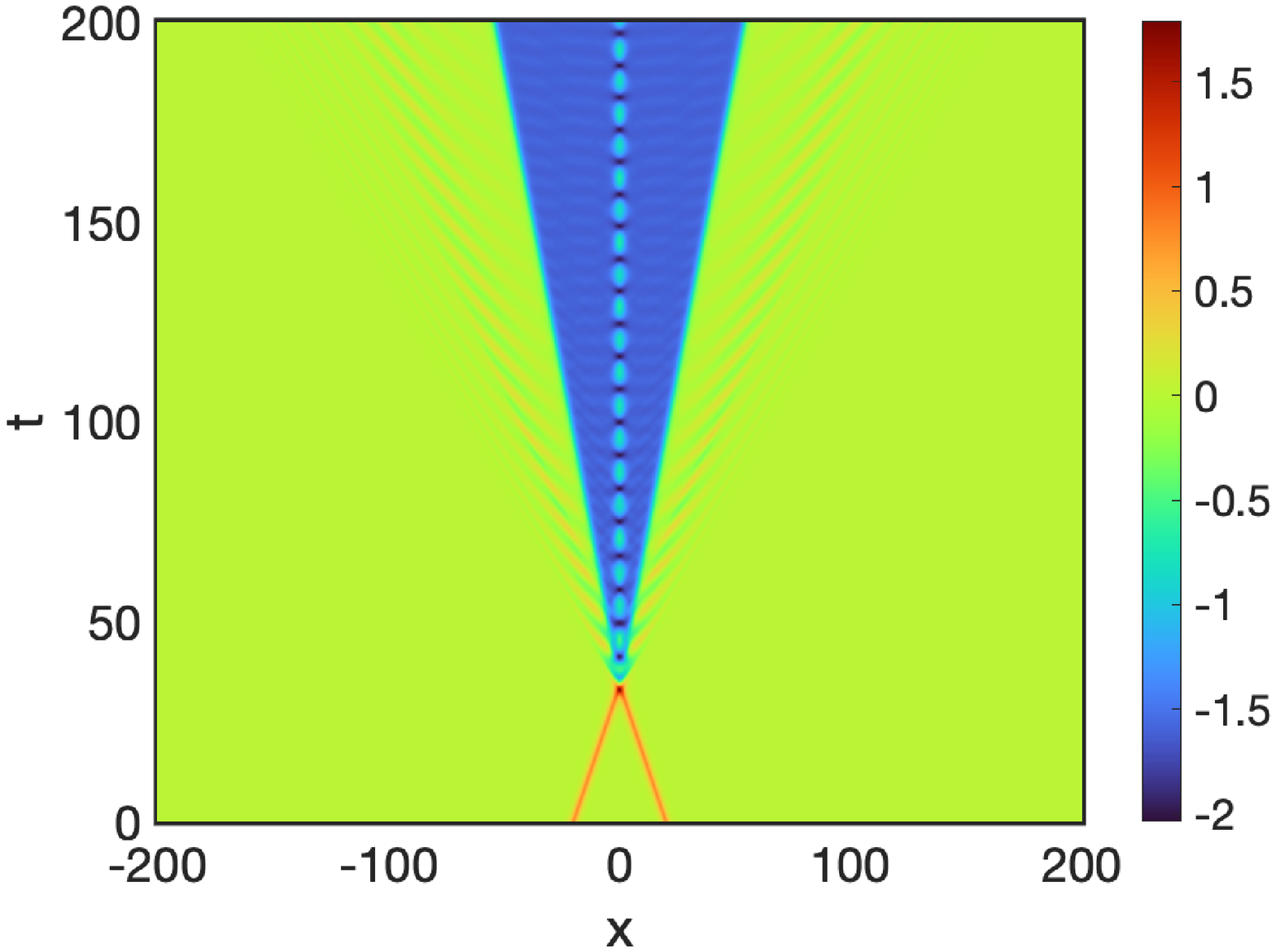}
	\caption{$\phi$-component (left) and $\chi$-component (right) for $v=0.586811$ with $r=0.391152$. }
\label{colz}
\end{figure}

We stress that we have not observed collisions with the characteristic of two-bounce resonance windows. The closest resemblance with a two-bounce window we observed for $0<r<1/2$ is depicted for instance in the Fig. \ref{colz}. Note that the kink-antikink pair collides twice. However, contrary to a two-bounce scattering, the original configuration is not recovered. Indeed, the vacuum at $x=0$ after the collision changes from $\phi=1$ to oscillations around $\phi=0$. The blue frontier between the I and D regions in the Fig. \ref {mosaic}b shows this outcome. We performed comprehensive numerical study at the frontiers of the ID, GI, HG, and FH zones but did not observe the emergence of two-bounce. In this way, despite the rich pattern of scattering, there is no evidence of a fractal structure similar to those of $n-$bounce resonance windows reported in the $\phi^4$ model and in the model discussed in the Ref. \cite{alonso7}.

%%%%%%%%%%%%%%%%%%%%%%%%%%%%%%%%%%%%%%%%%%%%%%%%%%%%%%%%%%%%%%%%%%%%%
%%%%%%%%%%%%%%%%%%%%%%%%%%%%%%%%%%%%%%%%%%%%%%%%%%%%%%%%%%%%%%%%%%%%%

%%%%%%%%%%%%%%%%%%%%%%%%%%%%%%%%%%%%%%%%%%%%%%%%%%%%%%%%%%%%%%%%%%%%%
%%%%%%%%%%%%%%%%%%%%%%%%%%%%%%%%%%%%%%%%%%%%%%%%%%%%%%%%%%%%%%%%%%%%%

%%%%%%%%%%%%%%%%%%%%%%%%%%%%%%%%%%%%%%%%%%%%%%%%%%%%%%%%%%%%%%%%%%%%%%%%%%

\section { Conclusions } \label{sec4}

%%%%%%%%%%%%%%%%%%%%%%%%%%%%%%%%%%%%%%%%%%%%%%%%%%%%%%%%%%%%%%%%%%%%%%%%%%
In this work we considered a two-scalar field model with kink solutions connecting five topological sectors, with four of them degenerate in energy. We considered kink-antikink scattering of the more energetic kink, $K_{21}$, from where an explicit $x$-dependence is known. We studied the effect of the coupling $r$ and initial velocity $v$ on the scattering. We noted that the phase diagram of the $\phi$-component matches that of the $\chi$-component. This shows that the scalar fields do not uncouple after the scattering, keeping the character of a nested defect. The structure of the phase diagram is complex, but it shows several regions where there the scattering is characterized by $K_{21} + \bar K_{21} \to K_{23} + \bar K_{23}$ or $K_{21} + \bar K_{21} \to K_{24} + \bar K_{24}$. There are differences in the regions concerned to the presence or absence of oscillations around $x=0$ and their degree of dispersion. There is a region where the collision results in complete annihilation of the pair, and another region where the collision results in transmutation to two thin $K_{24}\bar K_{24}$ pairs, both of which are manifestations that confirm the previously observed effects in Ref. \cite{alonso6}.

Despite the degeneracy in energy, the phase diagram is not symmetric concerning to the production of $K_{23}$ and $K_{24}$ kinks. This can be related to other aspects of these solutions, such as the presence of vibrational states from the linear perturbation analysis. However, for making a linear stability analysis, explicit solutions with dependence with $x$ and $r$ would be of interest. Unfortunately such solutions are not known for $K_{23}$ and $K_{24}$, in the whole range of values of $r$ which we consider in the present investigation. From the vacuum structure of the model, one would expect that $K_{23} \to K_{24}$ after the transformation $(\phi,\chi)\to(\phi,-\chi)$, a symmetry which is not present in the phase diagram displayed in the Fig. \ref{mosaic}. To better understand this issue, we note that the matrix potential of linear perturbations is not invariant under this transformation. To see how this works explicitly, let us  consider
\begin{eqnarray}
	\phi(x) &=& \phi_s(x) + \eta_n(x)\cos(\omega_n t),\\
	\chi(x) &=& \chi_s(x) + \xi_n(x)\cos(\omega_n t).
\end{eqnarray}
Substituting these equations into the equations of motion, we get the matrix operator 
\begin{eqnarray}
	\Bigg(\textbf{-1}\frac{d^2}{dx^2} + \textbf{M} \Bigg) \begin{pmatrix} \eta_n \\ \xi_n \end{pmatrix} = \omega^2_n \begin{pmatrix} \eta_n \\ \xi_n \end{pmatrix},
\end{eqnarray}
where the $\textbf{1}$ is the $2\times 2$ identity matrix and
\begin{eqnarray}
	\textbf{M} =  \begin{pmatrix} V_{\phi\phi} & V_{\phi\chi}  \\ V_{\chi\phi} & V_{\chi\chi} \end{pmatrix}
\end{eqnarray}
is the matrix potential of perturbations. For the model considered we have $V_{\phi\phi}= 6\phi^2 + (4r^2+2r)\chi^2-2$, $V_{\chi\chi}=(4r^2+2r)\phi^2+6r^2\chi^2-2r$ and $V_{\phi\chi}=V_{\chi\phi}=(8r^2+4r)\phi\chi$. That is, the non diagonal terms $V_{\phi\chi}$ and $V_{\chi\phi}$ of the matrix $M$ are not invariant under the transformation $(\phi,\chi)\to(\phi,-\chi)$. Then, despite symmetric and degenerate, the solutions $K_{23}$ and $K_{24}$ are not symmetric under linear perturbations, resulting in the complex behavior described in this work.

We also noted the absence of the generation of $K_{13}$ e $K_{14}$ in the scattering. This is due to the values of the vacuum: $(\phi,\chi)=(-1,0)$ at $x\to -\infty$ and $(\phi,\chi)=(1,0)$ at $x\to +\infty$. From the symmetry of the solutions, one expects to generate such kinks in the $\bar K_{21}  K_{21}$ scattering. Indeed, the pattern of the $\bar K_{21}  K_{21}$ scattering is the same observed for the $ K_{21}  \bar K_{21}$, with same output states after the transformation $K_{23} \to K_{13}$ and  $K_{24} \to K_{14}$.

%%%%%%%%%%%%%%%%%%%%%%%%%%%%%%%%%%%%%%%%%%%%%%%%%%%%%%%%%%%%%%%%%%%%%%%%%%%%%

\section{Acknowledgements}

F.C.S. and A.R.G. thank FAPEMA - Funda\c c\~ao de Amparo \`a Pesquisa e ao Desenvolvimento do Maranh\~ao through Grants PRONEM 01852/14, Universal 00920/19, 01191/16 and 01441/18. A.R.G. thanks CNPq (brazilian agency) through Grants 437923/2018-5 and 311501/2018-4 for financial support. This study was financed in part by the Coordena\c c\~ao de Aperfei\c coamento de Pessoal de N\'ivel Superior - Brasil (CAPES) - Finance Code 001. D.B. acknowledges CNPq (Grants No. 303469/2019-6 and No. 404913/2018-0) and Paraiba State Research Foundation (Grant 0015/2019) for financial support.

%%%%%%%%%%%%%%%%%%%%%%%%%%%%%%%%%%%%%%%%%%%%%%%%%%%%%%%%%%%%%%%%%%%%%%%%%%%%%

%%%%%%%%%%%%%%%%%%%%%%%%%%%%%%%%%%%%%%%%%%%%%%%%%%%%%%%%%%%%%%%%%%%%%%%%%%%%%%%

\end{document}